\documentclass[letterpaper]{article} 
\usepackage{aaai2026}  
\usepackage{times}  
\usepackage{helvet}  
\usepackage{courier}  
\usepackage[hyphens]{url}  
\usepackage{graphicx} 
\usepackage{natbib}  
\usepackage{caption} 
\usepackage{algorithm}

\usepackage{algpseudocode}

\newtheorem{proposition}{Propositions}

\usepackage[textsize=tiny]{todonotes}

\usepackage{subfig}
\usepackage{booktabs} 
\usepackage{amsmath}
\usepackage{amssymb}
\usepackage{cleveref}
\crefname{table}{Table}{Tables}
\crefname{appendix}{Appendix}{Appendixes}
\crefname{algorithm}{Algorithm}{Algorithms}
\crefname{figure}{Figure}{Figures}
\crefname{theorem}{Theorem}{}
\crefname{lemma}{Lemma}{}
\crefname{section}{Section}{}
\renewcommand{\eqref}[1]{Eq.~(\textup{\ref{#1}})}
\usepackage{multirow}
\usepackage{enumitem}
\usepackage{bm}

\usepackage{newfloat}
\usepackage{listings}
\DeclareCaptionStyle{ruled}{labelfont=normalfont,labelsep=colon,strut=off} 
\floatstyle{ruled}
\newfloat{listing}{tb}{lst}{}
\floatname{listing}{Listing}
\title{Hashed Watermark as a Filter: A Unified Defense Against Forging and Overwriting Attacks in Neural Network Watermarking}
\author{
    Yuan Yao\textsuperscript{\rm 1}, Jin Song\textsuperscript{\rm 2}, Jian Jin\textsuperscript{\rm 3}\thanks{Corresponding author.}
}
\affiliations{
    \textsuperscript{\rm 1} Beijing Teleinfo Technology Company Ltd., China Academy of Information and Communications Technology \\
    \textsuperscript{\rm 2} School of Computer Science, Nanjing University of Posts and Telecommunications \\
    \textsuperscript{\rm 3} Research Institute of Industrial Internet of Things, China Academy of Information and Communications Technology \\
    yaoyuan.hitsz@gmail.com, jinsongresearch@outlook.com, jin.jian@caict.ac.cn
}

\usepackage{bibentry}

\begin{document}

\maketitle

\begin{abstract}
As valuable digital assets, deep neural networks necessitate robust ownership protection, positioning neural network watermarking (NNW) as a promising solution.
Among NNW approaches, weight-based methods embed watermarks directly into model parameters; however, they remain generally susceptible to forging and overwriting attacks.
To address those challenges, we propose \textit{NeuralMark}, a robust method built around a \textit{hashed watermark filter}. 
Specifically, we utilize a hash function to generate an irreversible binary watermark from a secret key, which is then used as a filter to select the model parameters for embedding. 
This design cleverly intertwines the embedding parameters with the hashed watermark, providing a robust defense against both forging and overwriting attacks.
Average pooling is also incorporated to resist fine-tuning and pruning attacks.
Furthermore, it can be seamlessly integrated into various neural network architectures, ensuring broad applicability.
We theoretically analyze its security boundary and highlight the necessity of using a hashed watermark as a filter.
Empirically, we demonstrate its effectiveness and robustness across 13 distinct Convolutional and Transformer architectures, covering five image classification tasks and one text generation task. The source codes are available at \url{https://github.com/AIResearch-Group/NeuralMark}.
\end{abstract}


\section{Introduction}

The advancements in artificial intelligence have led to the development of numerous deep neural networks, particularly large language models \cite{mann2020language,achiam2023gpt,bai2023qwen,dubey2024llama,cao2024survey}. 
Training such models requires substantial investments in human resources, computational power, and other resources, as exemplified by GPT-4, which costs around \$$40$ million to train \cite{cottier2024rising}.
Neural networks can thus be regarded as valuable digital assets, making effective ownership protection essential. Motivated by this need, neural network watermarking (NNW) methods \cite{li2021survey,lukas2022sok,xue2021intellectual} have been proposed. They are generally categorized into three types: (i) White-box methods require access to the model’s internal information (\textit{e.g.}, parameters or activations) \cite{uchida2017embedding,liu2021watermarking,fan2021deepipr,li2024revisiting}; (ii) Black-box method require querying the model’s input–output mapping \cite{zhang2021deep,huang2023can,an2025decoder}; and (iii) Box-free method require only the model outputs and are particularly suitable for image generative models \cite{zhang2021deep,huang2023can,an2025decoder}. All three categories have demonstrated significant progress in safeguarding model ownership \cite{sun2023deep,ngo2025persistence}. Given the distinct challenges associated with each type, this work focuses on white-box NNW, leaving the investigation of other types for future work.

Existing white-box NNW methods can be broadly categorized into three sub-branches: (i) Weight-based methods \cite{uchida2017embedding,feng2020watermarking,li2021spread,liu2021watermarking,li2024revisiting} embed watermarks into model parameters; 
(ii) Passport-based methods \cite{fan2019rethinking,fan2021deepipr,zhang2020passport,liu2023trapdoor} introduce passport layers to replace normalization layers for watermark embedding; and (iii) Activation-based methods \cite{rouhani2019deepsigns,li2021feature,lim2022protect} incorporate watermarks into the activation maps of intermediate layers (see Appendix~A for a detailed discussion of related work). 
Among those methods, weight-based approaches embed watermarks directly into the model's parameters. This allows seamless integration into various network architectures without modifying the original structure \cite{uchida2017embedding,li2021survey}, providing a direct and easily implementable mechanism for watermark embedding.
Although several state-of-the-art weight-based methods \cite{feng2020watermarking,li2021spread,liu2021watermarking,li2024revisiting} can effectively resist fine-tuning and pruning attacks, \textit{they remain partially vulnerable to forging, overwriting, or both types of attacks}.

On the one hand, forging attacks attempt to fabricate counterfeit watermarks and infer the corresponding secret key through reverse engineering, by freezing the model parameters.
In this scenario, the adversary could claim the model's ownership, resulting in ownership ambiguity. 
On the other hand, overwriting attacks aim to remove the original watermark by embedding a counterfeit one. In particular, adversaries can adaptively increase the embedding strength of their watermarks without being required to match the original watermark's embedding strength. 
In such cases, the original watermark may be removed while the adversary's watermark is embedded, leading to the invalidation of the model's ownership. This raises a question: ``\textit{How can we design a more robust and effective weighted-based method that defends against both forging and overwriting attacks?}"

To explore this question, we propose \textit{NeuralMark}, a weighted-based method centered on a \textit{hashed watermark filter}.
Specifically, we use a hash function to generate an irreversible binary watermark from a secret key, which is then employed as a filter to select the model parameters for embedding. The \textit{avalanche effect} of the hash function \cite{webster1985design} ensures that slight input changes induce significant, unpredictable output variations, impeding gradient calculation and making reverse-engineering-based forging attacks infeasible. Moreover, using distinct hashed watermarks as private filters reduces parameter overlap, especially under repeated filtering, which increases the difficulty for adversaries to locate and manipulate the embedded parameters, thereby hindering overwriting attacks.
As a result, the hashed watermark filter cleverly intertwines the embedding parameters with the hashed watermark, providing a robust defense against both forging and overwriting attacks. 
Furthermore, we also apply an average pooling mechanism to the filtered parameters due to its resilience against fine-tuning and pruning attacks.
Upon obtaining the resulting parameters, the hashed watermark is embedded into those parameters using a lightweight embedding loss. 
During verification, the embedded watermark is extracted to verify model ownership.

The main contributions of this paper are threefold. 
\begin{itemize}
    \item We propose NeuralMark, a weight-based method designed to safeguard model ownership. Also, we provide a theoretical analysis of its security boundary.
    \item In NeuralMark, an elegant hashed watermark filter is developed to defend against both forging and overwriting attacks.
    \item Extensive experimental results across 13 distinct Convolutional and Transformer architectures, covering five image classification tasks and one text generation task, verify the effectiveness and robustness of NeuralMark.
\end{itemize}

\section{Threat Model}
\label{sec:threatModel}

In this section, we present the threat model considered in this work, detailing the adversary's capabilities and the corresponding success criteria.

\subsection{Adversary Capabilities}
We assume a \textit{fully trusted} third-party verifier responsible for watermark verification. An adversary can illegally access a watermarked model, identify the watermark-containing layers, and obtain the original training datasets, \textit{but is limited in computational resources}. This constraint is reasonable, as an attacker with sufficient computational resources could train a new model from scratch, making model theft unnecessary. As discussed above, this work focuses on forging and overwriting attacks, while also considering fine-tuning and pruning attacks. Those threat scenarios are detailed as follows.
(1) Forging Attack: The adversary aims to generate a counterfeit secret key–watermark pair without modifying the model parameters. Specifically, the adversary first randomly forges a counterfeit watermark and then derives a corresponding secret key by optimizing it while keeping the model parameters frozen \cite{fan2019rethinking,fan2021deepipr}.
(2) Overwriting Attack: The adversary attempts to embed a counterfeit watermark to overwrite the original one \cite{liu2021watermarking}. 
(3) Fine-tuning Attack: The adversary aims to fine-tune the model to remove the original watermark.
(4) Pruning Attack: The adversary attempts to remove the original watermark by parameter pruning.

\begin{figure*}[t]
\centering
\includegraphics[width=0.98\textwidth]{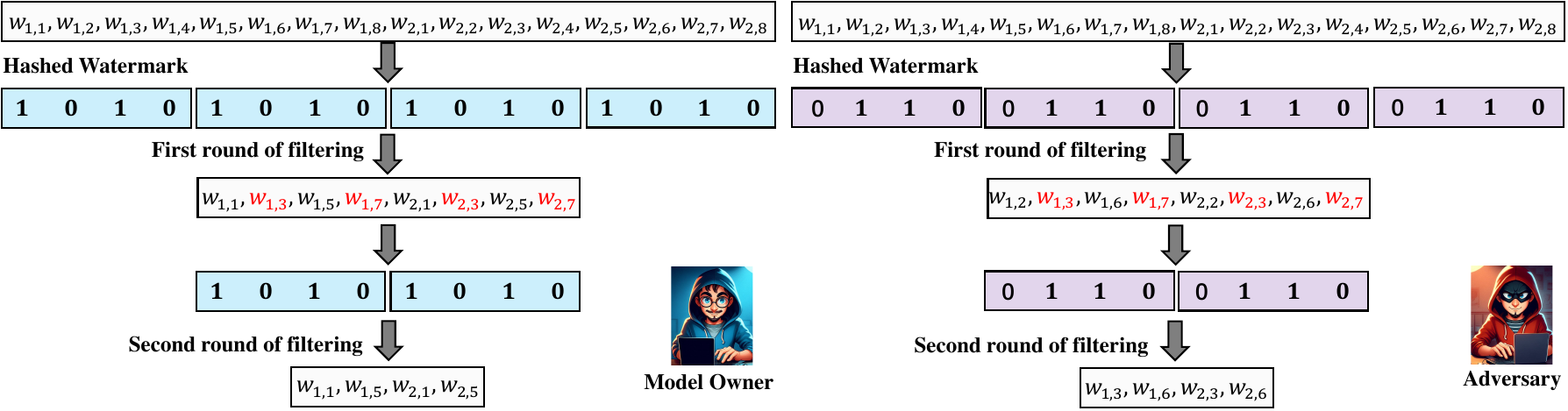}
\caption{Illustration of the hashed watermark filter. The model owner's hashed watermark is $[1, 0, 1, 0]$, while the adversary's is $[0, 1, 1, 0]$. The watermark is repeated to match the parameter length before each round of filtering. Without filtering, all 16 parameters overlap. After the first round, each watermark retains eight parameters with four overlapping; after the second round, only four parameters remain for each, with no overlap.}
\vspace{-1ex}
\label{fig:motivation}
\vspace{-2ex}
\end{figure*}

\subsection{Attack Success Criteria}
Building on insights from \cite{fan2019rethinking, fan2021deepipr, zhu2020secure, li2022leveraging}, a successful attack on a watermarked model typically requires the adversary to either (i) \textit{forge a counterfeit watermark without altering the model parameters}, or (ii) \textit{remove the original watermark through parameter modifications, all while preserving model performance}.
If the adversary only embeds a counterfeit watermark without removing the original one, the resulting model contains both. In this case, the model owner can submit a version containing only the original watermark to an authoritative third-party for verification. In contrast, the adversary cannot provide a model with only the counterfeit watermark, as the original watermark remains intact. As a result, the adversary cannot convincingly claim ownership unless they train a new model embedded solely with their own watermark. This not only makes stealing the original model unnecessary but also incurs significant training costs. 
Accordingly, we define the success criteria for each type of attack as follows.
(1) Success Criteria for Forging Attack: Forge a counterfeit watermark that passes verification without modifying the model parameters.
(2) Success Criteria for Overwriting Attack: Remove the original watermark and embed a counterfeit one by modifying the model parameters, while maintaining model performance.
(3) Success Criteria for Fine-tuning Attack: Remove the original watermark through fine-tuning, while maintaining model performance.
(4) Success Criteria for Pruning Attack: Remove the original watermark through parameter pruning, while maintaining model performance.

\section{Methodology}
\label{sec:methodology}

In this section, we present NeuralMark, a weight-based method designed to protect model ownership. The objective is to train a watermarked model $\mathbb{M}(\bm{\theta}^*)$ on a given training dataset $\mathcal{D}$ such that the model parameters $\bm{\theta}^*$ embed a binary watermark $\mathbf{b}$\footnote{Watermarks in this paper are binary vectors of 0s and 1s.} while satisfying the following criteria: (i) the watermark imposes negligible impact on the model performance and remains difficult for adversaries to detect; and (ii) the embedded watermark exhibits robustness against the adversarial attacks defined in the Threat Model section.

\subsection{Motivation}

As aforementioned, most weight-based methods struggle to defend against both forging and overwriting attacks. On the one hand, forging attacks aim to generate a counterfeit watermark and derive the corresponding secret key via gradient backpropagation, while keeping the model parameters fixed.
\textit{Defending against such attacks requires disrupting gradient computation to hinder reverse-engineering}. 
On the other hand, overwriting attacks attempt to remove the original watermark by embedding a counterfeit one. Once watermarked parameters are identified, the adversary can overwrite the original watermark.
Since each watermark updates the model parameters in a distinct and often conflicting direction, embedding a new watermark can easily disrupt the original one.
\textit{Defending against such attacks is essential to preserving the confidentiality of watermarked parameters and ensuring distinct parameter usage between the model owner and the adversary}.

To address both attacks, we propose a \textit{hashed watermark filter}, which uses an irreversible watermark generated from a secret key via a hash function as a private filter, restricting watermark embedding to a secret parameter subset.
This design provides two key properties: 
\begin{itemize}[left=0pt] 

\item \textbf{Gradient Obfuscation}: The avalanche effect of the hash function ensures that even minor input changes lead to large, unpredictable output variants, effectively impeding gradient computation and rendering reverse-engineering-based forging attacks infeasible.

\item \textbf{Embedding Isolation}: Since the hashed watermarks of the model owner and the adversary are inherently distinct, using them as private filters can effectively reduce the overlap in selected parameters, especially when the filtering process is performed repeatedly. As exemplified in \cref{fig:motivation}, the model owner's hashed watermark is $[1, 0, 1, 0]$, while the adversary's is $[0, 1, 1, 0]$. Without filtering, all 16 model parameters are shared, yielding a $100\%$ overlap ratio. After the first round of filtering, each party retains eight parameters, with four overlapping, reducing the overlap to $50\%$. A second filtering round results in four parameters per party, with zero overlap, achieving a $0\%$ overlap ratio. This progressive isolation ensures that as filtering continues, the overlap between the model owner’s and the adversary’s selected parameters is significantly reduced. Thus, it becomes increasingly difficult for the adversary to identify and manipulate the owner’s watermarked parameters, even when increasing the embedding strength of their watermarks, thereby preserving the integrity of the original watermark against overwriting attacks. 

\end{itemize}

In summary, these properties allow the hashed watermark filter to tightly entangle the embedding parameters with the hashed watermark, which is essential for resisting both forging and overwriting attacks (see the Security Analysis section for details). This mechanism forms the core of NeuralMark, which we will elaborate on next.

\subsection{NeuralMark}

NeuralMark consists of three primary steps: (i) hashed watermark generation; (ii) watermark embedding; and (iii) watermark verification. Figure 5 in Appendix C illustrates the workflow of each step. We now elaborate on each step.

\subsubsection{Hashed Watermark Generation}

As aforementioned, we construct a hash-based mapping from a secret key to a binary watermark. Formally, the watermark $\mathbf{b} \in \{0, 1\}^n$ is generated by $\mathbf{b} = \mathcal{H}(\mathbf{K})$, where $\mathbf{K} \in \mathbb{R}^{k \times n}$ is a secret key matrix with elements drawn from a random distribution (\textit{e.g.}, standard Gaussian distribution), $\mathcal{H}(\cdot)$ denotes a hash function, and $n$ indicates the length of the watermark.
To accommodate various application requirements, we adopt SHAKE-256 \cite{dworkin2015sha}, an extendable-output function from the SHA-3 family that allows dynamic adjustment of output length.  
Furthermore, auxiliary content $\mathcal{C}$ (\textit{e.g.}, textual descriptors or unique identifiers) can also be incorporated into the hash function, yielding $\mathbf{b} = \mathcal{H} (\mathbf{K} || \mathcal{C})$, where $||$ denotes the concatenation operation. This mechanism enables context-aware watermark generation without compromising the avalanche effect of the hash function.
For simplicity, we omit auxiliary content in the experiments.

\subsubsection{Watermark Embedding}

To embed the hashed watermark $\mathbf{b}$ into the model $\mathbb{M}(\bm{\theta})$, we first select and flatten a subset of parameters (\textit{e.g.}, one-layer parameters) from $\bm{\theta}$ into a parameter vector $\mathbf{w} \in \mathbb{R}^m$.
Then, we utilize the hashed watermark filter to select the model parameters for embedding. Specifically, let $\mathbf{w}^{(0)} = \mathbf{w}$ be the initial parameter vector. In the $r$-th ($r \in \{1, \cdots, R\}$) filtering round, the watermark $\mathbf{b}$ is repeated to match the length of $\mathbf{w}^{(r-1)}$, forming $\mathbf{b}^{(r)}$, with any excess parameters in $\mathbf{w}^{(r-1)}$ discarded. 
The parameter vector $\mathbf{w}^{(r)}$ is constructed by selecting the elements from $\mathbf{w}^{(r-1)}$ at positions where $\mathbf{b}^{(r)}$ equals one, \textit{i.e.}, $\mathbf{w}^{(r)} = \big[ w_i^{(r-1)} \mid i \in \{ j \mid b_j^{(r)} = 1 \} \big]$, where $w^{(r-1)}_i$ is the $i$-th element of $\mathbf{w}^{(r-1)}$, and $b^{(r)}_j$ is the $j$-th element of  $\mathbf{b}^{(r)}$. After completing the whole watermark filtering process, the filtered parameter vector $\mathbf{w}^{(R)}$ is obtained. 
Next, we adopt the average pooling $\text{AVG} (\cdot)$ operation \cite{gholamalinezhad2020pooling} to calculate the final parameters as $\widetilde{\mathbf{w}} = \text{AVG} (\mathbf{w}^{(R)}) \in \mathbb{R}^{k}$. This operation aggregates parameters across broader regions, thereby enhancing robustness against parameter perturbations caused by fine-tuning and pruning attacks.
Finally, we formulate the overall optimal objective as
\begin{equation}
    \label{loss:Total}
    \min_{\theta} \mathcal{L}_m + \lambda \mathcal{L}_e (\widetilde{\mathbf{b}}, \mathbf{b}),
    \end{equation}
where $\mathcal{L}_m$ denotes the main task loss (\textit{e.g.}, classification loss), $\mathcal{L}_e (\cdot, \cdot)$ represents the binary cross-entropy loss, $\widetilde{\mathbf{b}} = \delta(\widetilde{\mathbf{w}} \mathbf{K})$ denotes the extracted watermark, with $\delta(\cdot)$ being the sigmoid function, and $\lambda$ is a positive trade-off hyper-parameter.
By minimizing \eqref{loss:Total}, the watermark can be embedded into model parameters during the main task training. The watermark embedding process is summarized in Algorithm 1 in Appendix D.

\subsubsection{Watermark Verification}
The watermark verification process is similar to the embedding process. Concretely, upon identifying a potentially unauthorized model, the relevant subset of model parameters is extracted and subjected to hashed watermark filtering and average pooling to derive an extracted watermark $\widetilde{\mathbf{b}}$.
This extracted watermark is then compared to the model owner's watermark $\mathbf{b}$ using the \textit{watermark detection rate}, which is defined by
\begin{equation}
\label{eq:eta}
\rho = \frac{1}{n} \sum_{i = 1}^{n} \mathbf{1} \big[ b_i = \mathcal{T} (\widetilde{b}_i) \big],
\end{equation}
where $\mathcal{T}(x)$ is a threshold function that outputs $1$ if $x > 0.5$ and $0$ otherwise, and $\mathbf{1} (\psi)$ is an indicator function that returns 1 if $\psi$ is true and 0 otherwise. 
The unauthorized model is confirmed to belong to the model owner if both of the following conditions are satisfied: 
\textbf{(1)} The watermark detection rate $\rho$ exceeds a theoretical security boundary $\rho^\ast$, which will be theoretically analyzed later.
\textbf{(2)} The watermark must correspond to the output of the hash function applied to the secret key, ensuring cryptographic consistency with the predefined hash function.
The watermark verification process is outlined in Algorithm 2 in Appendix D.

\begin{table*}[t]
  \centering
  \caption{Comparison of classification accuracy (\%) across distinct datasets using AlexNet and ResNet-18. Watermark detection rates are omitted as they all reach 100\%.}
  \resizebox{\textwidth}{!}{
    \begin{tabular}{c|cc|cc|cc|cc|cc}
    \toprule
    \multirow{2}[4]{*}{Dataset} & \multicolumn{2}{c|}{Clean} & \multicolumn{2}{c|}{NeuralMark} & \multicolumn{2}{c|}{VanillaMark} & \multicolumn{2}{c|}{GreedyMark} & \multicolumn{2}{c}{VoteMark} \\
\cmidrule{2-11} & AlexNet & ResNet-18 & AlexNet & ResNet-18 & AlexNet & ResNet-18 & AlexNet & ResNet-18 & AlexNet & ResNet-18 \\
    \midrule
    CIFAR-10 & 91.05  & 94.76  & 90.93  & 94.50  & 91.01  & 94.87  & 90.88  & 94.69 & 90.86 & 94.79  \\
    CIFAR-100 & 68.24  & 76.23  & 68.57  & 76.34  & 68.43  & 76.22  & 68.31  & 76.14  & 68.53 & 76.74   \\
    Caltech-101 & 68.07  & 68.83  & 68.38  & 68.47  & 68.54  & 68.99  & 68.59  & 69.08  & 68.88 & 67.91  \\
    Caltech-256 & 44.27  & 54.09  & 44.55  & 53.71  & 44.73  & 53.47  & 44.64  & 53.28  & 44.43 & 54.71  \\
    TinyImageNet & 42.42  & 53.48  & 42.31  & 53.22  & 42.50  & 53.36  & 42.94  & 53.31  & 42.50 & 53.47  \\
    \bottomrule
    \end{tabular}%
    }
  \label{tab:fidelity_all_datasets}%
\end{table*}%
\begin{table*}[t]
  \centering
  \caption{Comparison of classification accuracy (\%) on CIFAR-100 using various architectures. Watermark detection rates are omitted as they all reach 100\%.}
  \resizebox{\textwidth}{!}{
    \begin{tabular}{c|cccccccccc}
    \toprule
     Method   & ViT-B/16 & Swin-V2-B & Swin-V2-S & VGG-16 & VGG-13 & ResNet-34 & WideResNet-50 & GoogLeNet & MobileNet-V3-L \\
    \midrule
    Clean & 39.07 & 52.99 & 55.88 & 72.75 & 72.71 & 77.06 & 59.67 & 60.71 & 61.11 \\
    NeuralMark & 39.22 & 53.57 & 55.87 & 72.61 & 71.49 & 77.03 & 58.41 & 60.02 & 61.8 \\
    \bottomrule
    \end{tabular}%
    }
  \label{tab:fidelity_all_architectures}%
\end{table*}%
\begin{table*}[!ht]
  \centering
  \caption{Comparison on E2E using GPT-2-S and GPT-2-M. Watermark detection rates are omitted as they all reach 100\%.}
    \resizebox{\textwidth}{!}{
    \begin{tabular}{c|ccccc|c|ccccc}
    \toprule
    GPT-2-S & BLEU  & NIST  & MET   & ROUGE-L & CIDEr & GPT-2-M & BLEU  & NIST  & MET   & ROUGE-L & CIDEr \\
    \midrule
    Clean & 69.36 & 8.76  & 46.06 & 70.85 & 2.48  & Clean & 68.7  & 8.69  & 46.38 & 71.19 & 2.5 \\
    NeuralMark & 69.59 & 8.79  & 46.01 & 70.85 & 2.48  & NeuralMark & 67.73 & 8.57  & 46.07 & 70.66 & 2.47 \\
    \bottomrule
    \end{tabular}%
    }
  \label{tab:fidelity_GPT}%
  \vspace{-1ex}
\end{table*}%

\subsection{Security Analysis}
\label{sec:throrem}

\subsubsection{Security Boundary Analysis}
We present a theoretical analysis to determine the security boundary of NeuralMark in Proposition 1.

\begin{proposition} 
\label{theo:Boundary} 
Under the assumption that the hash function produces uniformly distributed outputs \cite{bellare1993random}, for a model watermarked by NeuralMark with a watermark tuple $\{\mathbf{K}, \mathbf{b} \}$, where $\mathbf{b} = \mathcal{H} (\mathbf{K})$, if an adversary attempts to forge a counterfeit watermark tuple $\{\mathbf{K}^\prime, \mathbf{b}^\prime \}$ such that $\mathbf{b}^\prime = \mathcal{H} (\mathbf{K}^\prime)$ and $\mathbf{K}^\prime \neq \mathbf{K}$, then the probability of achieving a watermark detection rate of at least $\rho$ (i.e., \(\geq \rho\)) is upper-bounded by $\frac{1}{2^n} \sum_{i = 0}^{n - \lceil \rho n \rceil} \binom{n}{i}$.
\end{proposition}

The proof of Proposition 1 is provided in Appendix B. \Cref{theo:Boundary} provides a theoretical benchmark for establishing the security boundary of the watermark detection rate. Specifically, with $n = 256$, if the watermark detection rate $\rho \geq 88.29\%$, the probability of this occurring by forgery is less than $1 / 2^{128}$. This negligible probability allows us to confirm ownership with high confidence. \textit{Thus, we set $n = 256$ and use $88.29\%$ as the security bound for the watermark detection rate in the experiments}.

\subsubsection{Necessity of Hashed Watermark Filter}
We analyze the necessity of the hashed watermark filter by comparing it to a baseline mechanism that employs a private filter rather than a hashed watermark.
While this mechanism offers resistance to overwriting attacks, it remains vulnerable to forging attacks. 
For example, an adversary can use a $256 \times 256$ \textit{identity matrix} as a secret key $\mathbf{K}$ to generate a hashed watermark $\mathbf{b}$.
By selecting embedding parameters $\widehat{\mathbf{w}}$ whose signs correspond to $\mathbf{b}$ (with 0 representing a negative value and 1 a positive value),
the adversary can derive a private filter that selects those parameters accordingly.
This allows bypassing watermark verification, \textit{i.e.}, $\mathcal{T}(\delta(\widehat{\mathbf{w}} \mathbf{K})) = \mathbf{b}$ and $\mathcal{H} (\mathbf{K}) = \mathbf{b}$.
In contrast, the hashed watermark filter cleverly intertwines the embedding parameters with the hashed watermark, rendering it essential for defending against both forging and overwriting attacks.

\section{Experiments}
\label{sec:experiments}

In this section, we evaluate the proposed NeuralMark.

\subsection{Experimental Setup}

\subsubsection{Datasets and Architectures}

We use five image classification datasets: CIFAR-10 \cite{krizhevsky2009learning}, CIFAR-100 \cite{krizhevsky2009learning}, Caltech-101 \cite{fei2004learning}, Caltech-256 \cite{griffin2007caltech}, and TinyImageNet \cite{le2015tiny}, as well as one text generation dataset, E2E \cite{novikova2017e2e}. Additionally, we utilize 11 image classification architectures, including eight Convolutional architectures: AlexNet \cite{krizhevsky2012imagenet}, VGG-13, VGG-16 \cite{simonyan2014very}, GoogLeNet \cite{szegedy2015going}, ResNet-18, ResNet-34 \cite{he2016deep}, WideResNet-50 \cite{zagoruyko2016wide}, and MobileNet-V3-L \cite{howard2019searching}, as well as three Transformer architectures: ViT-B/16 \cite{dosovitskiy2020image}, Swin-V2-B, and Swin-V2-S \cite{liu2022swin}. Furthermore, we adopt two text generation architectures: GPT-2-S and GPT-2-M \cite{radford2019language}.

\subsubsection{Baselines and Metrics}

We compare NeuralMark with three popular weight-based methods presented in \cite{uchida2017embedding}, \cite{liu2021watermarking}, and \cite{li2024revisiting}, referred to as \textbf{VanillaMark}, \textbf{GreedyMark}, and \textbf{VoteMark}, respectively (see the Related Work section in Appendix A for details). Additionally, we include a comparison with a method that does not involve watermark embedding, referred to as \textbf{Clean}. For the image classification task, we assess model performance using classification accuracy, while the watermark embedding task is evaluated based on the watermark detection rate. 
As for the text generation task, we follow \cite{hu2022lora} and evaluate model performance using BLEU, NIST, MET, ROUGE-L, and CIDEr metrics, with the watermark embedding task assessed based on the watermark detection rate. 
More experimental details are provided in Appendix~E.

\subsection{Fidelity Evaluation}

\subsubsection{Diverse Datasets}
First, we evaluate the influence of watermark embedding on the model performance across diverse datasets.
\cref{tab:fidelity_all_datasets} reports the results across five image datasets using AlexNet and ResNet-18. 
We observe that all methods have minimal impact on model performance while successfully embedding watermarks, indicating that NeuralMark and other methods 
maintain model performance across diverse datasets during watermark embedding.

\subsubsection{Various Architectures}
Next, we assess the impact of NeuralMark on model performance across various architectures. \cref{{tab:fidelity_all_architectures}} lists the results of NeuralMark on the CIFAR-100 dataset using ViT-B/16, Swin-V2-B, Swin-V2-S, VGG-16, VGG-13, ResNet-34, WideResNet-50, GoogLeNet, and  MobileNet-V3-L.
We find that NeuralMark maintains a $100\%$ watermark detection rate across a wide range of architectures while exerting minimal impact on model performance. Those observations indicate that NeuralMark exhibits a good level of generalizability across architectures.

\subsubsection{Text Generation Tasks}
Finally, we evaluate the effect of NeuralMark on the text generation tasks. \cref{tab:fidelity_GPT} presents the results of NeuralMark applied to the GPT-2-S and GPT-2-M architectures on the E2E dataset. We can observe that NeuralMark achieves a $100\%$ watermark detection rate while maintaining nearly lossless model performance. Those results demonstrate NeuralMark’s potential and generality in ownership protection of text generative models.

\begin{table}[H]
  \centering
  \caption{Comparison of detection rate (\%) of \textit{\textbf{counterfeit}} watermarks using ResNet-18.}
  \resizebox{\columnwidth}{!}{
    \begin{tabular}{c|cccc}
    \toprule
    Dataset & NeuralMark & VanillaMark & GreedyMark & VoteMark \\
    \midrule
    CIFAR-10 & 48.56  & 100.00  & 50.70  & 100.00  \\
    CIFAR-100 & 49.41 & 100.00  & 52.85  & 100.00  \\
    \bottomrule
    \end{tabular}%
    }
  \label{tab:VanillaForging}%
\end{table}%
\begin{table*}[t]
  \centering
  \caption{Comparison of resistance to overwriting attacks at various trade-off hyper-parameters ($\lambda$) and learning rates ($\eta$) using ResNet-18. Values (\%) \textbf{inside} and outside the bracket are \textbf{watermark detection rate} and classification accuracy, respectively. Adversary watermarks, which are consistently detected at 100\%, are omitted.}
  \resizebox{\textwidth}{!}{
    \begin{tabular}{c@{\hskip 3pt}|c@{\hskip 3pt}|c@{\hskip 3pt}c@{\hskip 3pt}c@{\hskip 3pt}c@{\hskip 3pt}|c@{\hskip 3pt}|c@{\hskip 3pt}c@{\hskip 3pt}c@{\hskip 3pt}c}
    \toprule
    Overwriting & $\lambda$ & NeuralMark & VanillaMark & GreedyMark & VoteMark & $\eta$ & NeuralMark & VanillaMark & GreedyMark & VoteMark \\
    \midrule
    \multirow{5}[2]{*}{\parbox{2cm}{CIFAR-100 \\ \centering to \\CIFAR-10}} & 1 & 93.65 (\textbf{100}) & 93.30 (\textbf{100}) & 93.45 (\textbf{48.82}) & 93.63 (\textbf{100}) & 0.001 & 93.65 (\textbf{100}) & 93.30 (\textbf{100}) & 93.45 (\textbf{48.82}) & 93.63 (\textbf{100}) \\
          & 10    & 93.44 (\textbf{100}) & 93.58 (\textbf{100}) & 93.29 (\textbf{51.17}) & 93.13 (\textbf{100}) & 0.005 & 91.76 (\textbf{99.60}) & 92.17 (\textbf{73.04}) & 92.13 (\textbf{50.00}) & 92.45 (\textbf{78.90}) \\
          & 50    & 93.46 (\textbf{100}) & 93.50 (\textbf{100}) & 93.07 (\textbf{55.07}) & 93.39 (\textbf{100}) & 0.01  & 91.58 (\textbf{92.18}) & 91.79 (\textbf{62.10}) & 91.53 (\textbf{49.60}) & 91.76 (\textbf{60.15}) \\
          & 100   & 93.53 (\textbf{100}) & 92.95 (\textbf{94.53}) & 93.18 (\textbf{54.29}) & 93.53 (\textbf{96.48}) & 0.1   & 75.2 (\textbf{50.78}) & 79.68 (\textbf{47.26}) & 72.42 (\textbf{53.12}) & 70.92 (\textbf{54.29}) \\
          & 1000  & 93.09 (\textbf{100}) & 92.89 (\textbf{53.90}) & 92.85 (\textbf{49.60}) & 92.77 (\textbf{59.37}) & 1     & 10.00 (\textbf{44.53}) & 10.00 (\textbf{53.51}) & 10.00 (\textbf{48.04}) & 10.00 (\textbf{53.51}) \\
    \bottomrule
    \end{tabular}%
    }
  \label{tab:OverwritingAttack_100to10}%
\end{table*}%
\begin{table*}[t]
  \centering
  \caption{Comparison of resistance to fine-tuning attacks using ResNet-18. Values (\%) \textbf{inside} and outside the bracket are \textbf{watermark detection rate} and classification accuracy, respectively.}
  \resizebox{\textwidth}{!}{
    \begin{tabular}{c@{\hskip 3pt}|c@{\hskip 3pt}c@{\hskip 3pt}|c@{\hskip 3pt}c@{\hskip 3pt}|c@{\hskip 3pt}c@{\hskip 3pt}|c@{\hskip 3pt}c@{\hskip 3pt}|c@{\hskip 3pt}c}
    \toprule
    \multirow{2}[4]{*}{Fine-tuning} & \multicolumn{2}{c|}{Clean} & \multicolumn{2}{c|}{NeuralMark} & \multicolumn{2}{c|}{VanillaMark} & \multicolumn{2}{c|}{GreedyMark} & \multicolumn{2}{c}{VoteMark} \\
\cmidrule{2-11}          & AlexNet & ResNet-18 & AlexNet & ResNet-18 & AlexNet & ResNet-18 & AlexNet & ResNet-18 & AlexNet & ResNet-18 \\
    \midrule
    CIFAR-100 to CIFAR-10 & 85.55 & 89.15 & 85.35(\textbf{100}) & 88.83(\textbf{100}) & 85.48(\textbf{91.01}) & 89.35(\textbf{85.93}) & 80.41(\textbf{96.48}) & 76.15(\textbf{94.14}) & 84.97(\textbf{89.06}) & 89.66(\textbf{85.54}) \\
    CIFAR-10 to CIFAR-100 & 58.96 & 49.74 & 58.50(\textbf{100}) & 49.77(\textbf{100}) & 58.75(\textbf{74.21}) & 49.97(\textbf{70.31}) & 51.75(\textbf{97.65}) & 19.94(\textbf{82.42}) & 58.81(\textbf{80.07}) & 49.08(\textbf{71.87}) \\
    Caltech-256 to Caltech-101 & 47.65 & 74.09 & 71.29(\textbf{100}) & 73.12(\textbf{100}) & 71.56(\textbf{100}) & 74.03(\textbf{100}) & 72.04(\textbf{100}) & 68.45(\textbf{100}) & 71.62(\textbf{100}) & 72.47(\textbf{99.60}) \\
    Caltech-101 to Caltech-256 & 40.61 & 40.00  & 40.34(\textbf{100}) & 40.34(\textbf{100}) & 40.71(\textbf{96.09}) & 39.04(\textbf{93.36}) & 40.68(\textbf{100}) & 36.45(\textbf{98.82}) & 39.52(\textbf{95.31}) & 39.73(\textbf{93.75}) \\
    \bottomrule
    \end{tabular}%
    }
  \label{tab:fine-tuningWatermarkLayer}%
\end{table*}%

\subsection{Robustness Evaluation}

\subsubsection{Forging Attacks}

We follow the setting described in the Threat Model section to evaluate the robustness of NeuralMark against forging attacks.
Specifically, for VanillaMark and VoteMark, we randomly generate a counterfeit watermark and then attempt to learn the corresponding secret key while keeping the model parameters fixed.
As for GreedyMark and NeuralMark, we directly verify 10 randomly forged watermarks using the watermarked model because GreedyMark does not require a secret key, and NeuralMark benefits from the avalanche effect of the hash function and the tight coupling between the embedding parameters and the hashed watermark, making reverse-engineering infeasible.
\cref{tab:VanillaForging} presents the detection rates of counterfeit watermarks, from which we draw the following observations. 
(1) For VanillaMark and VoteMark, a pair of counterfeited secret key and watermark can be successfully learned through reverse-engineering, indicating their vulnerability to forging attacks.
(2) NeuralMark and GreedyMark demonstrate robust resistance against forging attacks, which aligns with our expectations.

\subsubsection{Overwriting Attacks}

We conduct overwriting attacks targeting the watermark embedding layers, with the number of training epochs fixed at 100 to reflect limited computational resources. The optimization is guided by the loss function $\mathcal{L}_m + \lambda \mathcal{L}_e(\widetilde{\mathbf{b}}, \mathbf{b}_a)$, where $\mathbf{b}_a$ denotes the adversary's watermark.
Also, we analyze the effects of two key factors: the hyperparameter $\lambda$ and the learning rate $\eta$. Here, $\lambda$ controls the strength of the watermark embedding, with larger values leading to stronger embedding, while $\eta$ primarily affects model performance.

\noindent\textbf{Distinct Values of $\bm{\lambda}$}. We investigate the influence of $\lambda$ in overwriting attacks. 
Specifically, we set $\lambda$ to 1, 10, 50, 100, and 1000, respectively. \cref{tab:OverwritingAttack_100to10} presents the results on the CIFAR-100 to CIFAR-10 task using ResNet-18.
We report only the original watermark detection rate, as the adversary's watermark detection rate reaches $100\%$. 
As defined in the success criterion in the Threat Model section, the original watermark must be effectively removed for overwriting attacks to be deemed successful. Thus, the overwriting attack experiments focus solely on whether the original watermark can be successfully removed.
We can summarize several insightful observations. 
\textbf{(1)} As $\lambda$ increases, the original watermark detection rate of NeuralMark remains at 100\%, while those of VanillaMark, GreedyMark, and VoteMark significantly decline. In particular, when $\lambda = 1000$, the embedding strength of the adversary's watermark is 1000 times greater than that of the original watermark. At this point, the original watermark detection rates for NeuralMark, VanillaMark, GreedyMark, and VoteMark on the CIFAR-100 to CIFAR-10 task are $100\%$, $53.90\%$, $49.60\%$, and $59.37\%$, respectively. Those results indicate that NeuralMark exhibits strong robustness against overwriting attacks.
\textbf{(2)} As $\lambda$ increases, model performance remains relatively stable. This is because overwriting attacks jointly train both the main task and the watermark embedding task, enabling the model parameters to effectively adapt to both. More results are offered in Appendix F.1.

\noindent\textbf{Distinct Values of $\bm{\eta}$}. We examine the impact of $\eta$ in overwriting attacks. Concretely, we set $\eta$ to 0.001, 0.005, 0.01, 0.1, and 1, respectively. \cref{tab:OverwritingAttack_100to10} lists the results on the CIFAR-100 to CIFAR-10 task using ResNet-18. 
We have several important observations.
\textbf{(1)} Larger $\eta$ values hurt model performance, implying that the adversary cannot arbitrarily increase the attack strength.
\textbf{(2)} At $\eta = 0.005$, the original watermark detection rates for VanillaMark, GreedyMark, and VoteMark drop sharply, whereas NeuralMark maintains a detection rate close to $100\%$. 
\textbf{(3)} When $\eta = 0.01$, the model performance of NeuralMark on the CIFAR-100 to CIFAR-10 task decreases by $2.07\%$, but its original watermark detection rate remains above the security boundary of 88.29\% defined in the Security Boundary Analysis section, while those for the other methods fall significantly. 
\textbf{(4)} For $\eta >= 0.1$, although the original watermark detection rate of NeuralMark drops below the security boundary, the model performance is completely compromised, indicating that the attack is ineffective. More results are provided in Appendix F.1.

\subsubsection{Fine-tuning Attacks}
We perform fine-tuning attacks on the watermark embedding layers. During the attack, the task-specific classifier is first replaced with randomly initialized parameters, after which only the parameters of the watermark embedding layers and the classifier are updated, while all other parameters remain frozen.
The optimization is guided solely by the main task loss $\mathcal{L}_m$.
Following \cite{liu2021watermarking}, we adopt the same hyper-parameters for fine-tuning attacks as during training, except for setting the learning rate to 0.001. As shown in \cref{tab:fine-tuningWatermarkLayer}, 
we find that watermarks embedded with NeuralMark maintain a $100\%$ watermark detection rate across all fine-tuning tasks. In contrast, watermarks embedded with VanillaMark, GreedyMark, and VoteMark experience a slight reduction in detection rates across several tasks. 
Those results indicate that fine-tuning attacks cannot effectively remove watermarks embedded with NeuralMark. Furthermore, we conduct a fine-tuning attack by updating all model parameters, as detailed in Appendix~F.2.

\subsubsection{Pruning Attacks}
\begin{figure}[htbp]
\centering
\subfloat[NeuralMark] 
{
{\includegraphics[width=0.23\textwidth]{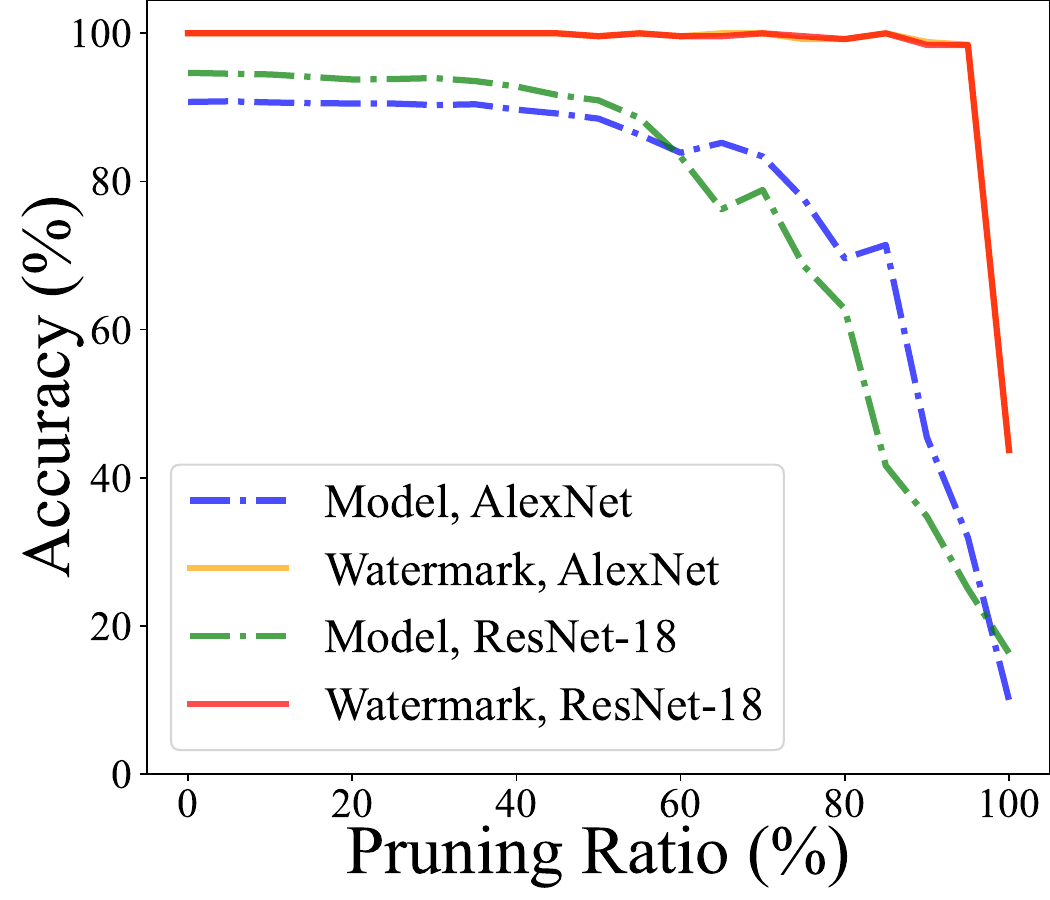}}
}
\hspace{-2.5mm}
\subfloat[VanillaMark] 
{
{\includegraphics[width=0.23\textwidth]{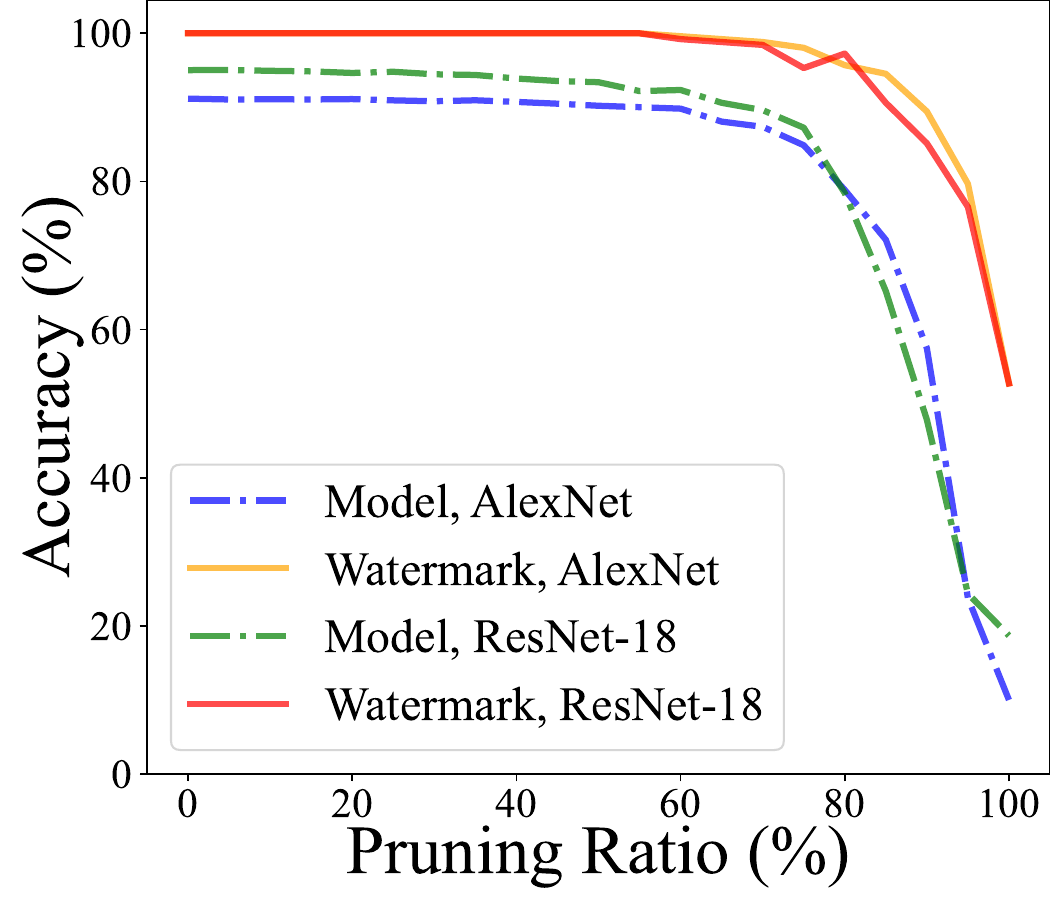}}
}
\caption{Comparison of resistance to pruning attacks under various pruning ratios on CIFAR-10 using AlexNet and ResNet-18.}
\label{fig:PruningAttackCIFAR-10Main}
\vspace{-2ex}
\end{figure}

We evaluate the robustness of NeuralMark against pruning attacks by randomly resetting a specified proportion of parameters in the watermark embedding layer to zero. \cref{fig:PruningAttackCIFAR-10Main} presents the results of NeuralMark and VanillaMark on the CIFAR-10 dataset using AlexNet and ResNet-18, respectively. As the pruning ratio increases, NeuralMark's performance degrades slightly, while the detection rate remains nearly 100\%, indicating a good level of robustness. Additional results for all baselines across different datasets are provided in Appendix F.3.

\subsection{Analysis}
\label{sec:analysis}

\subsubsection{Parameter Distribution}

\cref{fig:DistributionResNet-18} shows the parameter distributions learned by Clean and NeuralMark on the CIFAR-100 dataset using ResNet-18. As observed, their distributions are nearly indistinguishable, making it difficult for adversaries to detect the embedded watermarks. Additional results across various architectures are provided in Appendix F.4.

\subsubsection{Performance Convergence}

\cref{fig:ConvergenceResNet-18} shows the performance convergence of Clean and NeuralMark on the CIFAR-100 dataset using ResNet-18. The two curves follow a similar trajectory and remain closely aligned, indicating that NeuralMark does not hinder model convergence. Additional results across various architectures are provided in Appendix F.5.

\begin{figure}[t]
    \centering
    \subfloat[Distribution \label{fig:DistributionResNet-18}]{%
        \includegraphics[width=0.22\textwidth]{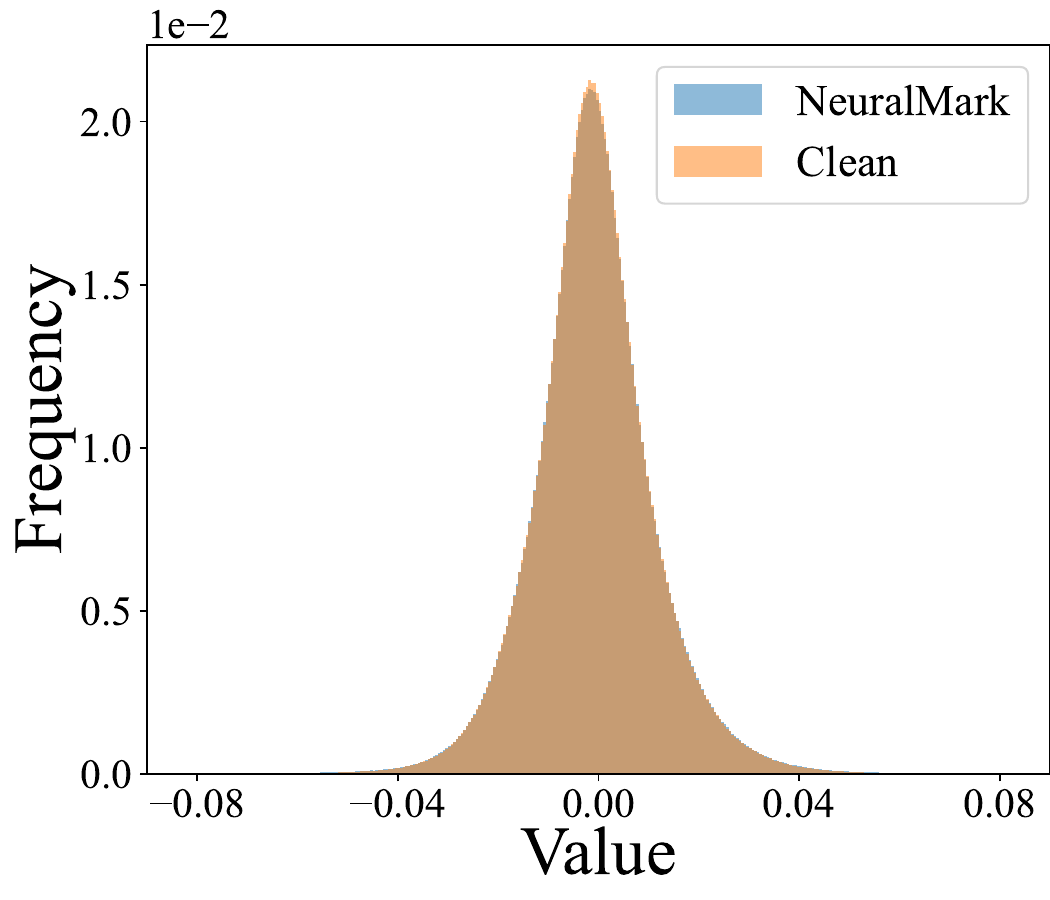}
    }
    \hspace{-1.5mm}
    \subfloat[Convergence \label{fig:ConvergenceResNet-18}]{%
        \includegraphics[width=0.21\textwidth]{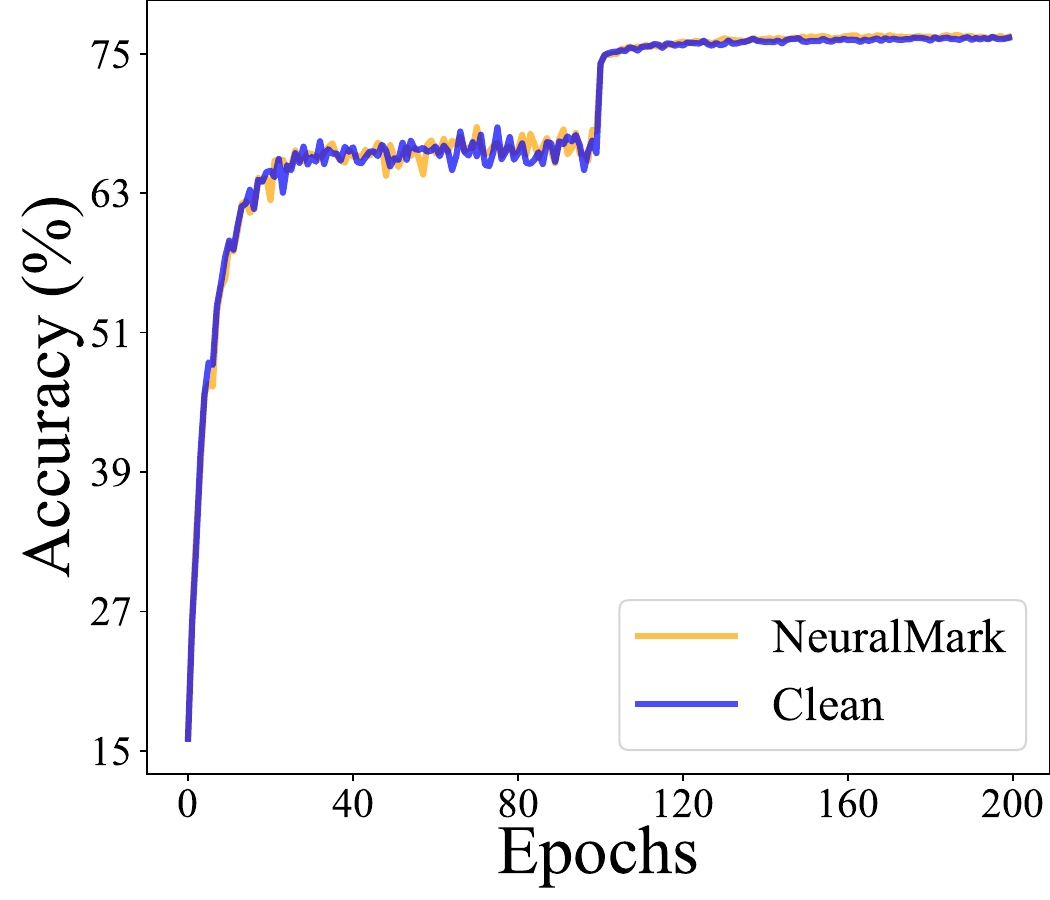}
    }
    \caption{Parameter distribution and performance convergence on the CIFAR-100 dataset using ResNet-18.}
    \label{fig:DistributionConvergence}
    \vspace{-2ex}
\end{figure}

\subsubsection{Filtering Rounds} 
\begin{figure}[htbp]
\centering
\includegraphics[width=0.25\textwidth]{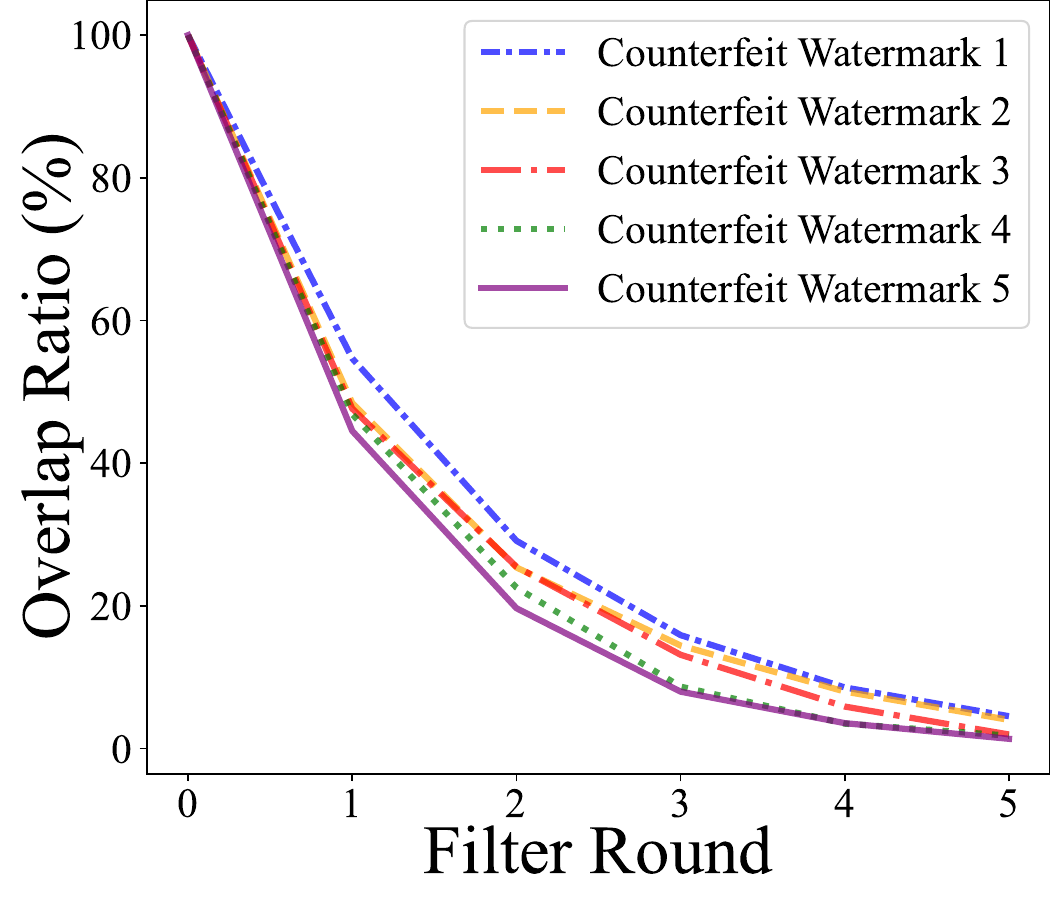}
\caption{Comparison of parameter overlap ratio with different filter rounds on CIFAR-100 using ResNet-18.}
\label{fig:filter}
\vspace{-2ex}
\end{figure}

To analyze watermark filtering efficacy, we generate five counterfeit watermarks and calculate the overlap ratio between parameters filtered with those and the original watermark. As shown in \cref{fig:filter}, the overlap rate decreases towards zero with more filtering rounds, indicating that watermark filtering enhances the secrecy of the watermarked parameters. 
Furthermore, Appendix G presents additional experiments with 6 and 8 filtering rounds to evaluate their impact on NeuralMark’s effectiveness robustness against various attacks, compared to the default setting of 4. The results show that the number of filtering rounds has a negligible effect on robustness.

\subsubsection{Additional Analyses} 
The impact of the watermark embedding layers and watermark length on model performance, as well as the training efficiency, is analyzed in Appendices F.6–F.8, respectively. Those results demonstrate the flexibility, effectiveness, and efficiency of NeuralMark.

\section{Conclusion}
\label{sec:conclusion}

In this paper, we present NeuralMark, a white-box method designed to protect model ownership. At the core of NeuralMark is a hashed watermark filter, which utilizes a hash function to generate an irreversible binary watermark from a secret key, subsequently employing this watermark as a filter to select model parameters for embedding. 
We provide a theoretical analysis of its security boundary and highlight the necessity of employing a hashed watermark as a filter. 
Extensive experiments on various datasets, architectures, and tasks confirm NeuralMark's effectiveness and robustness. 
In future work, we plan to investigate how the proposed hashed watermark filter can be incorporated with existing watermarking approaches to offer complementary protection against broader attack scenarios.

\bibliography{NeuralMark}

\newpage
\onecolumn
\appendix

We provide additional details and experimental results in the appendices. Below are the contents.

\begin{itemize}

\item Appendix A: Related Work.

\item Appendix B: Proof of Proposition 1.

\item Appendix C: Workflow of NeuralMark.

\item Appendix D: Algorithms of NeuralMark.

\item Appendix E: Implementation Details. 

\item Appendix F: Additional Experimental Results.

\item Appendix G: Further Analysis on Filtering Rounds.

\end{itemize}

\section{A. Related Work}
\label{sec:relatedwork}

In this section, we review weight-based, passport-based, and activation-based methods, respectively.

\noindent\textbf{Weight-based Method}. This kind of methods \cite{uchida2017embedding,feng2020watermarking,li2021spread,liu2021watermarking,li2024revisiting} embeds watermarks into the model parameters of neural networks.
For instance, \cite{uchida2017embedding} proposes the first weight-based method, which embeds the watermark into the model parameters of an intermediate layer in the neural network. 
Another example is \cite{li2021spread}, which presents a method based on spread transform dither modulation that enhances the secrecy of the watermark.
However, those two methods cannot effectively resist forging and overwriting attacks. Moreover, \cite{feng2020watermarking} utilizes the secret keys to pseudo-randomly select parameters for watermark embedding and apply spread-spectrum modulation to disperse the modulated watermark across different layers.
This method effectively defends against overwriting attacks while neglecting forging attacks. 
Additionally, \cite{liu2021watermarking} proposes to greedily choose important model parameters for watermark embedding without an additional secret key. Although this method is effective against forging attacks, it fails to provide strong resistance to overwriting attacks of varying strength levels. 
Recently, \cite{li2024revisiting} introduces random noises into the watermarked parameters and then employs a majority voting scheme to aggregate the verification results across multiple rounds. While this method enhances the watermark's robustness to some extent, it remains ineffective against forging and overwriting attacks.

\noindent\textbf{Passport-based Method}. This group of methods \cite{fan2019rethinking,fan2021deepipr,zhang2020passport,liu2023trapdoor} 
integrates the watermark into the normalization layers in neural networks. Specifically, \cite{fan2019rethinking,fan2021deepipr} propose the first passport-based method, which utilizes additional passport samples (\textit{e.g.}, images) to generate affine transformation parameters for the normalization layers, tightly binding them to the model performance. 
Subsequently, \cite{zhang2020passport} integrates a private passport-aware branch into the normalization layers, which is trained jointly with the target model and is used solely for watermark verification. 
Recently, \cite{liu2023trapdoor} argues that binding the model performance is insufficient to defend against forging attacks, and thus proposes establishing a hash mapping between passport samples and watermarks.

\noindent\textbf{Activation-based Method}. This category of methods \cite{rouhani2019deepsigns,li2021feature,lim2022protect} incorporates watermarks into the activation maps of intermediate layers in neural networks. For instance, \cite{rouhani2019deepsigns} incorporates the watermark into the mean vector of activation maps generated by predetermined trigger samples. Similarly, \cite{li2021feature} directly integrates the watermark into the activation maps associated with the trigger samples. Additionally, \cite{lim2022protect} embeds the watermark into the hidden memory state of a recurrent neural network.

\section{B. Proof for Proposition 1}
\label{appendix:ProofforBoundary}


\textbf{Proposition 1.}
\textit{Under the assumption that the hash function produces uniformly distributed outputs \cite{bellare1993random}, for a model watermarked by NeuralMark with a watermark tuple $\{\mathbf{K}, \mathbf{b} \}$, where $\mathbf{b} = \mathcal{H} (\mathbf{K})$, if an adversary attempts to forge a counterfeit watermark tuple $\{\mathbf{K}^\prime, \mathbf{b}^\prime \}$ such that $\mathbf{b}^\prime = \mathcal{H} (\mathbf{K}^\prime)$ and $\mathbf{K}^\prime \neq \mathbf{K}$, then the probability of achieving a watermark detection rate of at least $\rho$ (i.e., \(\geq \rho\)) is upper-bounded by $\frac{1}{2^n} \sum_{i = 0}^{n - \lceil \rho n \rceil} \binom{n}{i}$.}

\textit{Proof}. Since the hash function produces uniformly distributed outputs, each bit of the counterfeit watermark matches the corresponding bit of the extracted watermark from model parameters with a probability of $\frac{1}{2}$. The number of matching bits follows a binomial distribution with parameters $n$ and $p = \frac{1}{2}$. To achieve a detection rate of at least 
$\rho$, the adversary needs at least $\lceil \rho n \rceil$ bits to match out of $n$ bits. Thus, the probability of having at least \(\lceil \rho n \rceil\) matching bits is given by
\begin{equation}
\begin{split}
   \Pr \big[X \geq \lceil \rho n \rceil \big] & = \sum_{i = \lceil \rho n \rceil}^{n} \binom{n}{i} \left(\frac{1}{2}\right)^i \left(\frac{1}{2}\right)^{n-i}
   \\ &
   = \frac{1}{2^n} \sum_{i = \lceil \rho n \rceil}^{n} \binom{n}{i} = \frac{1}{2^n} \sum_{i = 0}^{n - \lceil \rho n \rceil} \binom{n}{i}.
\end{split}
\end{equation}
Accordingly, the probability of an adversary forging a counterfeit watermark that achieves a watermark detection rate of at least \(\rho\) (\textit{i.e.}, \(\geq \rho\)) is upper-bounded by $\frac{1}{2^n} \sum_{i = 0}^{n - \lceil \rho n \rceil} \binom{n}{i}$.

\section{C. Workflow of NeuralMark}

\begin{figure*}[htbp]
\centering
\subfloat[Generation]{
    \includegraphics[width=0.167\textwidth]{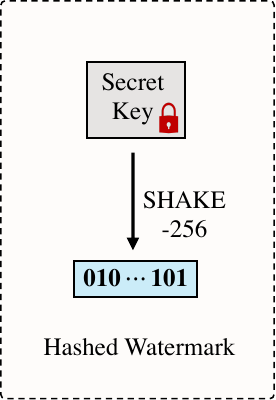}
    \label{fig:generation}
}
\hspace{-1ex}
\subfloat[Embedding]{
    \includegraphics[width=0.41\textwidth]{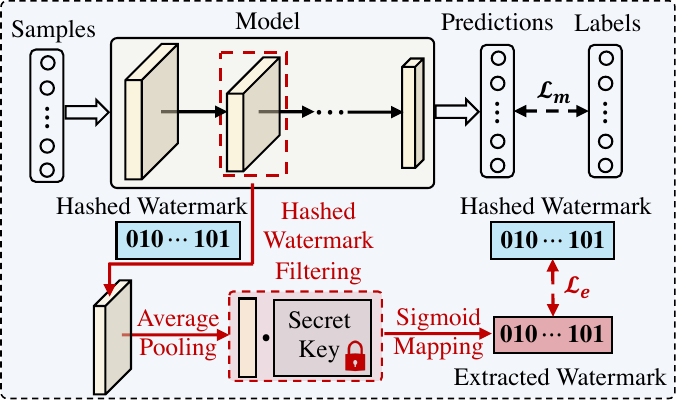}
    \label{fig:embedding}
}
\hspace{-1ex}
\subfloat[Verification]{
    \includegraphics[width=0.382\textwidth]{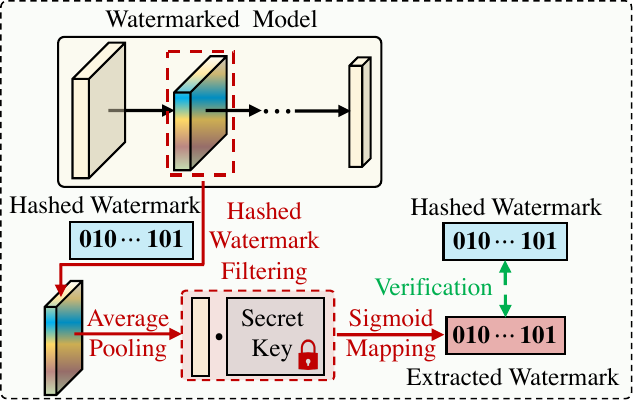}
    \label{fig:verification}
}
\caption{Illustrations of the processes for watermark generation (a), embedding (b), and verification (c).}
\label{appendix_fig:NeuralMark}
\end{figure*}

\cref{appendix_fig:NeuralMark} illustrates the workflow of NeuralMark, including watermark generation, embedding, and verification stages.

\section{D. Algorithms of NeuralMark}
\label{appendix:algorithm}

Algorithms~\ref{alg:embedding}-\ref{alg:verification} offer the watermark embedding and verification processes in NeuralMark, respectively.

\begin{algorithm}[htbp]
    \caption{Watermark Embedding in NeuralMark}
    \label{alg:embedding}
    \begin{algorithmic}[1]
    \Require Training dataset $\mathcal{D}$, secret key $\mathbf{K}$, index of embedding layer $\mathbf{I}_e$, hyper-parameters $\lambda$, $T$, and filter rounds $R$.
    \Ensure Watermarked model $\mathbb{M} (\theta^*)$.
    \State Randomly initialize the model parameter $\theta$.
    \State Generate the watermark $\mathbf{b} = \mathcal{H}(\mathbf{K})$.
    \For{$t = 0$ to $T - 1$}
        \State Use $\mathbf{I}_e$ to select a subset from $\theta$ and flatten it into $\mathbf{w}$.
         \For{$r = 1$ to $R$}
            \State Perform watermark filtering on $\mathbf{w}$ to obtain $\mathbf{w}^{(r)}$.
         \EndFor
        \State Apply average pooling on $\mathbf{w}^{(R)}$ to yield $\widetilde{\mathbf{w}}$.
        \State Execute sigmoid mapping on $\widetilde{\mathbf{w}} \mathbf{K}$ to produce $\widetilde{\mathbf{b}}$.
        \State Update $\theta$ based on \eqref{loss:Total}.
    \EndFor
     \end{algorithmic}
\end{algorithm}
\begin{algorithm}[htbp]
    \caption{Watermark Verification in NeuralMark}
    \label{alg:verification}
    \begin{algorithmic}[1]
    \Require Watermarked model $\mathbb{M} (\theta^*)$, secret key $\mathbf{K}$, watermark $\mathbf{b}$, index of embedding layer $\mathbf{I}_e$, filter rounds $R$, and security boundary $\rho^\ast$. 
    \Ensure True (Verification Success) or False (Verification Failure).
    \State Use $\mathbf{I}_e$ to select a subset from $\theta^*$ and flatten it to create $\mathbf{w}$.
    \For{$r = 1$ to $R$}
        \State Perform watermark filtering on $\mathbf{w}$ to obtain $\mathbf{w}^{(r)}$.
    \EndFor
    \State Apply average pooling on $\mathbf{w}^{(R)}$ to yield $\widetilde{\mathbf{w}}$.
    \State Execute sigmoid mapping on $\widetilde{\mathbf{w}} \mathbf{K}$ to produce $\widetilde{\mathbf{b}}$.
    \State Calculate watermark detection rate $\rho$ based on \eqref{eq:eta}.
    \If{$\rho \geq \rho^\ast$ \textbf{and} $\mathcal{H} (\mathbf{K}) = \mathbf{b}$}
        \State \textbf{return} True
    \Else
        \State \textbf{return} False
    \EndIf
    \end{algorithmic}
\end{algorithm}

\section{E. Implementation Details}

We implement NeuralMark using the PyTorch framework \cite{paszke2019pytorch} and conduct all experiments on three NVIDIA V100 series GPUs. The specific hyper-parameters are summarized below.
\begin{itemize}[left= 0 pt]
    \item For all the image classification architectures, we train for 200 epochs with a multi-step learning rate schedule from scratch, with learning rates set to $0.01$, $0.001$, and $0.0001$ for epochs 1 to 100, 101 to 150, and 151 to 200, respectively. We apply a weight decay of $5 \times 10^{-4}$ and set the momentum to $0.9$. The batch sizes for the training and test datasets are set to 64 and 128, respectively. In addition, we set hyper-parameter $\lambda$ to 1 and the number of filter rounds $R$ to 4. 
    
    \item For the GPT-2-S and GPT-2-M architectures, we use the Low-Rank Adaptation (LoRA) technique \cite{hu2022lora}. Each architecture is trained for 5 epochs with a linear learning rate scheduler, starting at $2 \times 10^{-4}$. We set the warm-up steps to 500, apply a weight decay with a coefficient of 0.01, and enable bias correction in the AdamW optimizer \cite{loshchilov2017fixing}. The dimension and the scaling factor for LoRA are set to 4 and 32, respectively, with a dropout probability of $0.1$ for the LoRA layers. The batch sizes for the training and test sets are 8 and 4, respectively. Moreover, we set hyper-parameter $\lambda$ to 1 and the number of filter rounds $R$ to 10.
\end{itemize}

\section{F. Additional Experimental Results}

\subsection{F.1 Overwriting Attacks}
\label{appendix:additionaloverwriting}

\cref{tab:OverwritingAttack} lists the results of the overwriting attack on the CIFAR-10 to CIFAR-100 task using ResNet-18, which are consistent with those observed on the CIFAR-100 to CIFAR-10 task reported in the main text. Those results further demonstrate that NeuralMark exhibits strong robustness against overwriting attacks across a range of attack strengths.

\begin{table}[htbp]
  \centering
  \caption{Comparison of resistance to overwriting attacks at various trade-off hyper-parameters ($\lambda$) and learning rates ($\eta$) using ResNet-18. Values (\%) \textbf{inside} and outside the bracket are \textbf{watermark detection rate} and classification accuracy, respectively.}
  \resizebox{\textwidth}{!}{
    \begin{tabular}{c|c|cccc|c|cccc}
    \toprule
    Overwriting & $\lambda$ & NeuralMark & VanillaMark & GreedyMark & VoteMark & $\eta$ & NeuralMark & VanillaMark & GreedyMark & VoteMark \\
    \midrule
    \multirow{5}[2]{*}{\parbox{2cm}{CIFAR-10 \\ \centering to \\CIFAR-100}} & 1     & 71.78 (\textbf{100}) & 72.68 (\textbf{98.82}) & 71.34 (\textbf{55.07}) & 72.97 (\textbf{98.43}) & 0.001 & 71.78 (\textbf{100}) & 72.68 (\textbf{98.82}) & 71.34 (\textbf{55.07}) & 72.97 (\textbf{98.43}) \\
          & 10    & 72.6 (\textbf{100}) & 72.03 (\textbf{98.04}) & 72.30 (\textbf{49.21}) & 72.08 (\textbf{98.04}) & 0.005 & 71.04 (\textbf{99.60}) & 70.02 (\textbf{69.53}) & 70.25 (\textbf{48.04}) & 71.11 (\textbf{71.09}) \\
          & 50    & 72.73 (\textbf{100}) & 72.45 (\textbf{95.70}) & 70.92 (\textbf{46.87}) & 72.38 (\textbf{97.26}) & 0.01  & 69.14 (\textbf{96.48}) & 69.02 (\textbf{59.76}) & 69.25 (\textbf{46.09}) & 68.88 (\textbf{62.11}) \\
          & 100   & 71.49 (\textbf{100}) & 71.92 (\textbf{92.18}) & 72.05 (\textbf{48.04}) & 72.72 (\textbf{93.75}) & 0.1   & 51.88 (\textbf{60.54}) & 51.76 (\textbf{53.90}) & 51.71 (\textbf{51.56}) & 51.74 (\textbf{56.25}) \\
          & 1000  & 71.81 (\textbf{100}) & 71.35 (\textbf{57.42}) & 71.74 (\textbf{51.95}) & 70.73 (\textbf{56.64}) & 1     & 1.00 (\textbf{44.53}) & 1.00 (\textbf{53.15}) & 1.00 (\textbf{50.00}) & 1.00 (\textbf{53.51}) \\
    \bottomrule
    \end{tabular}%
    }
  \label{tab:OverwritingAttack}%
\end{table}%

\subsection{F.2 Fine-tuning Attacks on All Model Parameters}

We conduct a fine-tuning attack by updating all model parameters. The results are reported in \cref{appendix_tab:FinetuningAttackAll}. As shown, NeuralMark consistently demonstrates superior robustness, indicating that fine-tuning attacks cannot effectively remove watermarks embedded with NeuralMark. Notably, the model performance of all methods improves substantially under this setting. Specifically, on the CIFAR-10 to CIFAR-100 task using ResNet-18, NeuralMark achieves an accuracy of 71.67\%, significantly higher than the 49.77\% accuracy obtained when only the watermark embedding layers and classifier are fine-tuned (see \cref{tab:fine-tuningWatermarkLayer}). Similar trends are observed across other methods. Those results suggest that fine-tuning only the watermark-specific layers and classifier makes it difficult to preserve model performance.

Furthermore, to assess the impact of the learning rate $\eta$ in fine-tuning attacks, we evaluate settings of $\eta = 0.01$ and $0.1$, in comparison to the default value of $0.001$. Those attacks are performed by updating all model parameters, allowing the model to retain effective model performance and thus posing a more practical threat. \cref{appendix_tab:additionalFine-tuning} presents the results on the CIFAR-100 to CIFAR-10 task using ResNet-18. At $\eta = 0.01$, NeuralMark consistently achieves a higher watermark detection rate than other methods, demonstrating its robustness. At $\eta = 0.1$, the model performance of all methods drops significantly, indicating that overly aggressive fine-tuning disrupts model functionality and renders the attack ineffective.

\begin{table}[htbp]
  \centering
  \caption{Comparison of resistance to fine-tuning attacks against all layers. Values (\%) \textbf{inside} and outside the bracket are \textbf{watermark detection rate} and classification accuracy, respectively.}
  \resizebox{\textwidth}{!}{
    \begin{tabular}{c@{\hskip 3pt}|c@{\hskip 3pt}c@{\hskip 3pt}|c@{\hskip 3pt}c@{\hskip 3pt}|c@{\hskip 3pt}c@{\hskip 3pt}|c@{\hskip 3pt}c@{\hskip 3pt}|c@{\hskip 3pt}c}
    \toprule
    \multirow{2}[4]{*}{Fine-tuning} & \multicolumn{2}{c|}{Clean} & \multicolumn{2}{c|}{NeuralMark} & \multicolumn{2}{c|}{VanillaMark} & \multicolumn{2}{c|}{GreedyMark} & \multicolumn{2}{c}{VoteMark} \\
\cmidrule{2-11}          & AlexNet & ResNet-18 & AlexNet & ResNet-18 & AlexNet & ResNet-18 & AlexNet & ResNet-18 & AlexNet & ResNet-18 \\
    \midrule
    CIFAR-100 to CIFAR-10 & 89.44 & 93.21 & 89.11(\textbf{100}) & 93.74(\textbf{100}) & 89.00(\textbf{100}) & 93.29(\textbf{100}) & 89.34(\textbf{99.21}) & 93.21(\textbf{100}) & 89.03(\textbf{100}) & 93.59(\textbf{100}) \\
    CIFAR-10 to CIFAR-100 & 65.46 & 72.17 & 64.60(\textbf{100}) & 71.67(\textbf{100}) & 65.03(\textbf{92.18}) & 72.49(\textbf{97.26}) & 64.57(\textbf{98.82}) & 72.06(\textbf{100}) & 64.83(\textbf{96.09}) & 72.27(\textbf{98.04}) \\
    Caltech-256 to Caltech-101 & 72.69 & 76.93 & 73.55(\textbf{100}) & 76.60(\textbf{100}) & 72.90(\textbf{100}) & 78.48(\textbf{100}) & 73.12(\textbf{100}) & 77.19(\textbf{100}) & 72.90(\textbf{100}) & 77.41(\textbf{100}) \\
    Caltech-101 to Caltech-256 & 43.39 & 46.48 & 43.15(\textbf{100}) & 44.42(\textbf{100}) & 43.21(\textbf{98.43}) & 45.69(\textbf{99.60}) & 43.47(\textbf{99.60}) & 45.25(\textbf{100}) & 43.78(\textbf{98.43}) & 45.29(\textbf{100}) \\
    \bottomrule
    \end{tabular}%
    }
  \label{appendix_tab:FinetuningAttackAll}%
\end{table}%
\begin{table}[htbp]
  \centering
  \caption{Comparison of resistance to fine-tuning attacks against all layers at various learning rates ($\eta$) using ResNet-18. Values (\%) inside and outside the bracket are the watermark detection rate and classification accuracy, respectively.}
    \begin{tabular}{c|c|ccccc}
    \toprule
    Fine-tuning & $\eta$  & Clean & NeuralMark & VanillaMark & GreedyMark & VoteMark \\
    \midrule
    \multirow{3}[2]{*}{CIFAR-100 to CIFAR-10} & 0.001 & 93.21  & 93.74(100) & 93.29(100) & 93.21(100) & 93.59(100) \\
          & 0.01  & 91.22  & 90.41(97.65) & 92.04(58.59) & 90.91(82.81) & 88.95(50.39) \\
          & 0.1   & 69.05  & 78.71(58.20) & 76.43(51.17) & 82.34(62.10) & 84.72(50.00) \\
    \bottomrule
    \end{tabular}%
  \label{appendix_tab:additionalFine-tuning}%
  \vspace{-1ex}
\end{table}%

\subsection{F.3 Pruning Attacks}
\label{appendix:additionalpruning}

\begin{figure}[htbp]
\centering
\subfloat[NeuralMark] 
{
{\includegraphics[width=0.24\textwidth]{./Figures/Pruning/NeuralMark/cifar10.pdf}}
}
\hspace{-2.5mm}
\subfloat[VanillaMark] 
{
{\includegraphics[width=0.24\textwidth]{./Figures/Pruning/VanillaMark/cifar10.pdf}}
}
\hspace{-2.5mm}
\subfloat[GreedyMark]
{
{\includegraphics[width=0.24\textwidth]{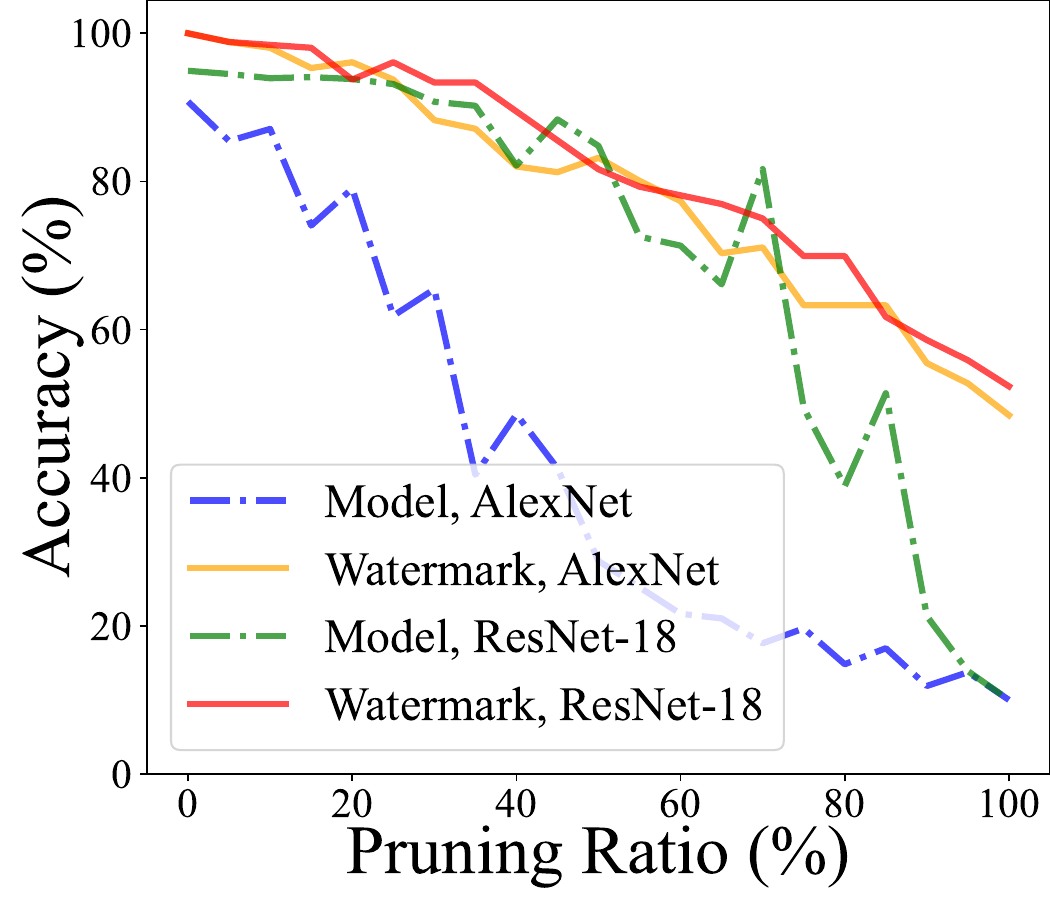}}
}
\hspace{-2.5mm}
\subfloat[VoteMark]
{
{\includegraphics[width=0.24\textwidth]{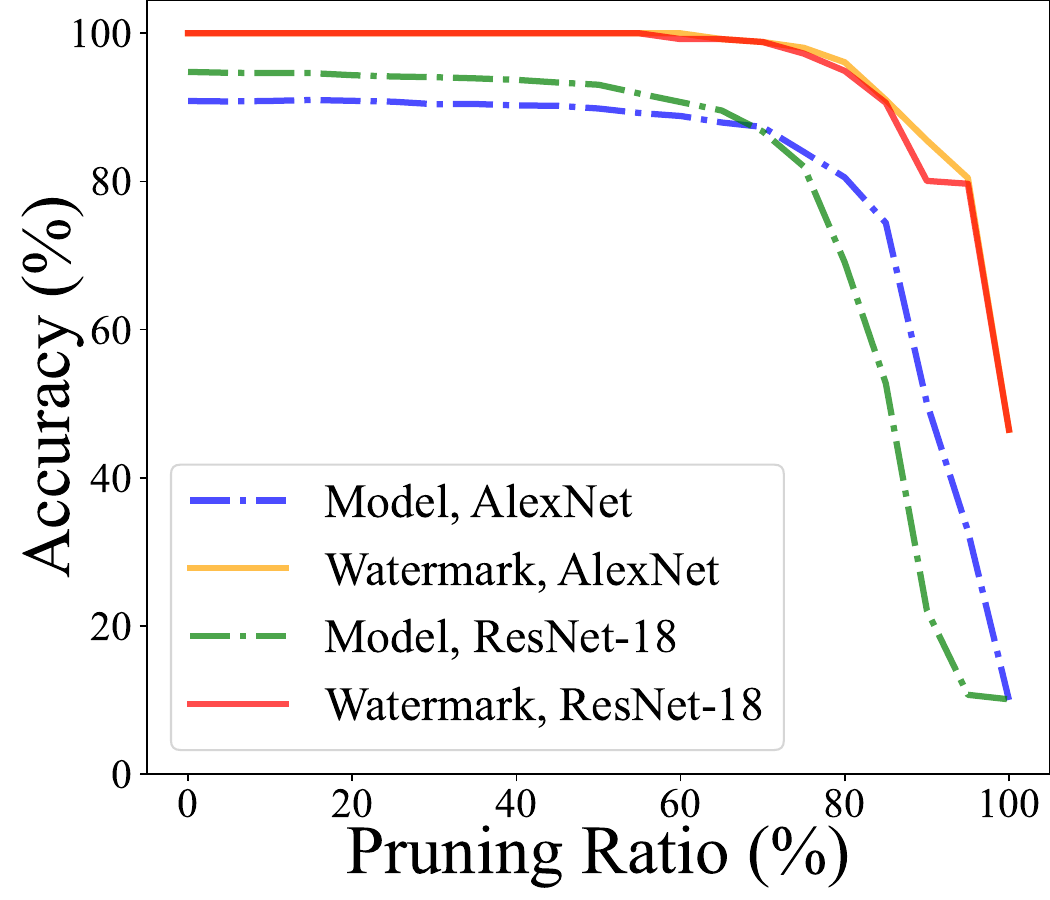}}
}
\caption{Comparison of resistance to pruning attacks under various pruning ratios on CIFAR-10 using AlexNet and ResNet-18.}
\label{appendix_fig:PruningAttackCIFAR-10}
\vspace{-1ex}
\end{figure}

\begin{figure}[htbp]
\vspace{-1ex}
\centering
\subfloat[NeuralMark] 
{
{\includegraphics[width=0.22\textwidth]{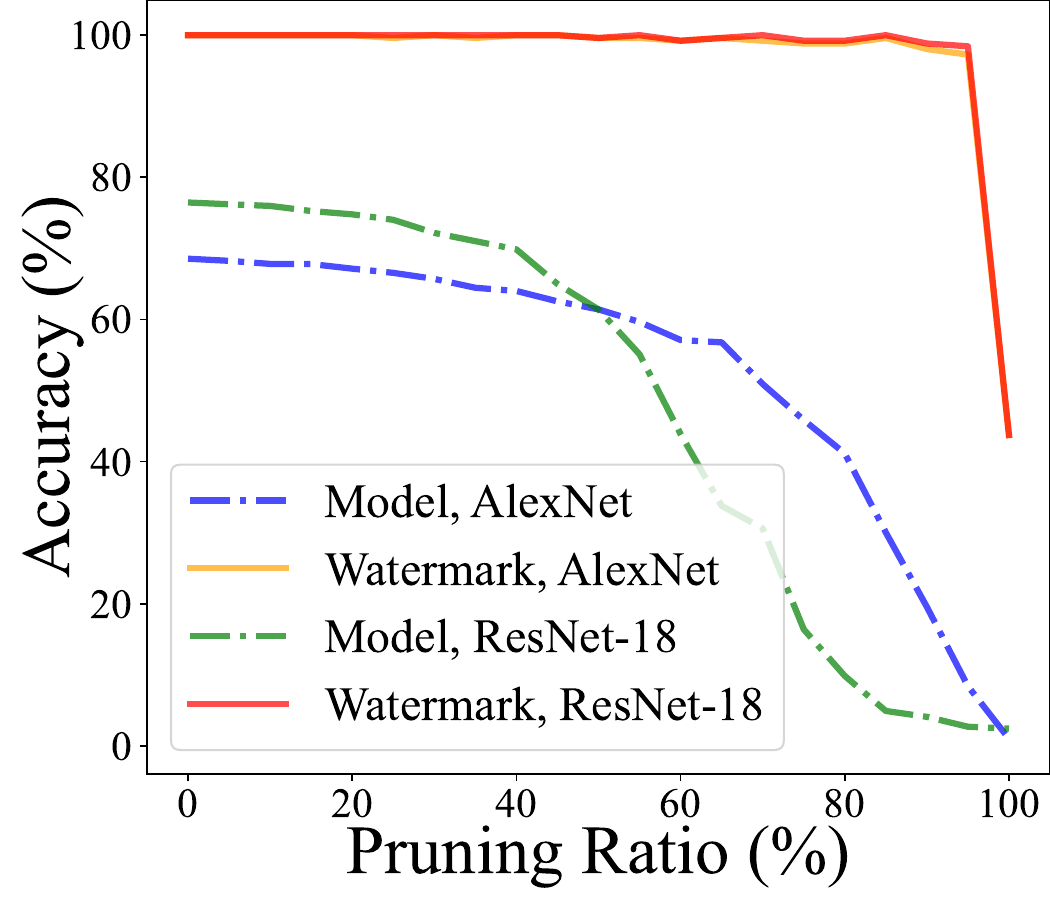}}
}
\hspace{-2.5mm}
\subfloat[VanillaMark] 
{
{\includegraphics[width=0.22\textwidth]{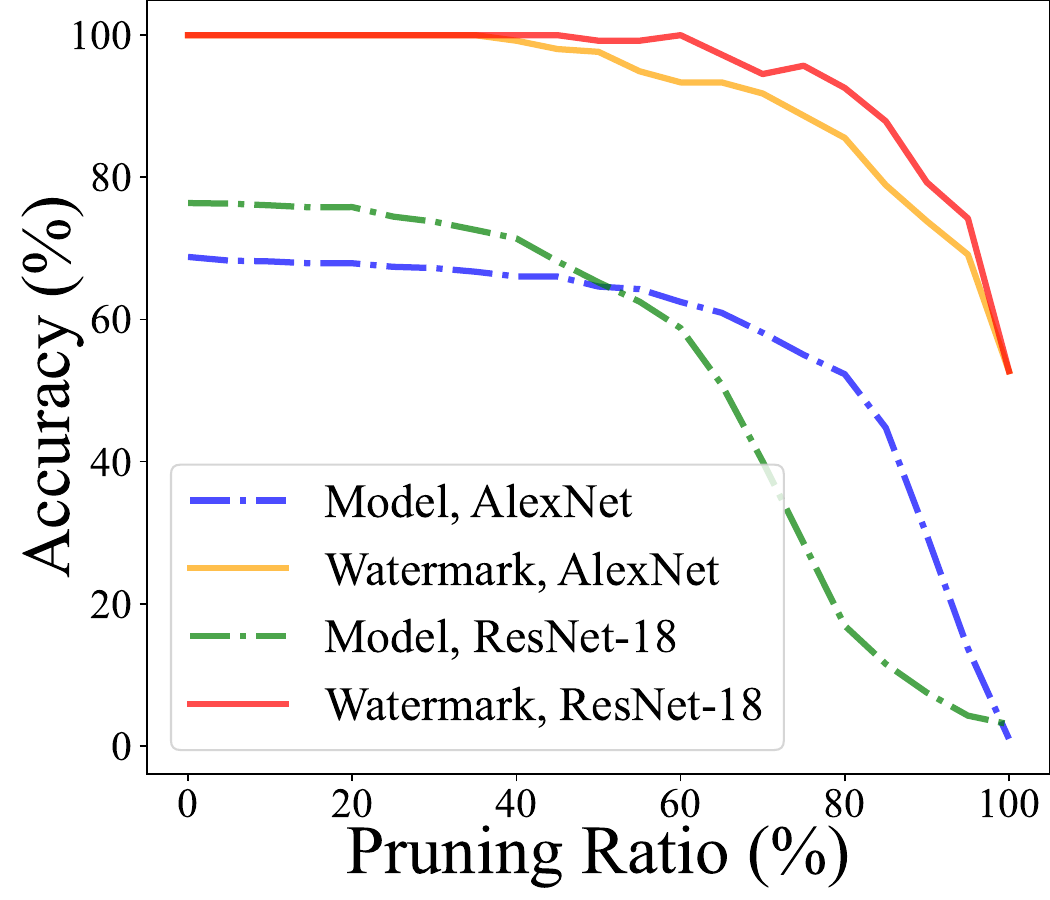}}
}
\hspace{-2.5mm}
\subfloat[GreedyMark]
{
{\includegraphics[width=0.22\textwidth]{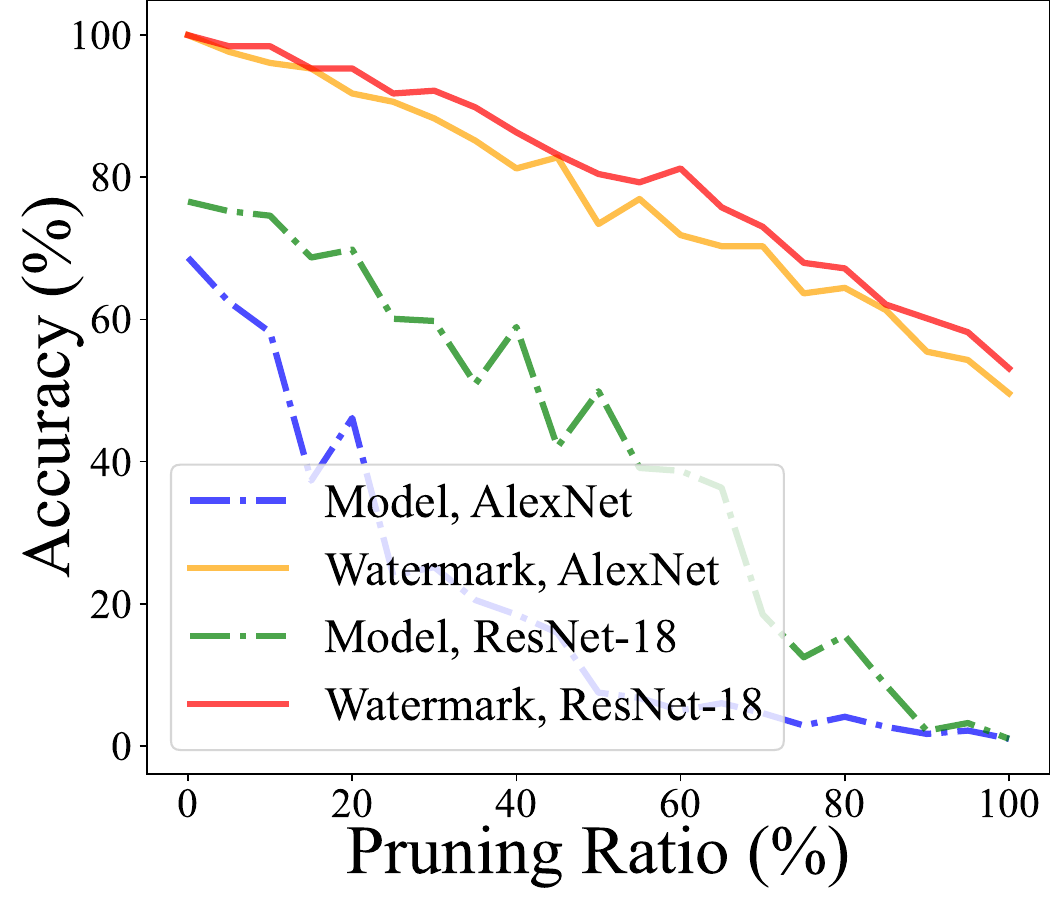}}
}
\hspace{-2.5mm}
\subfloat[VoteMark]
{
{\includegraphics[width=0.22\textwidth]{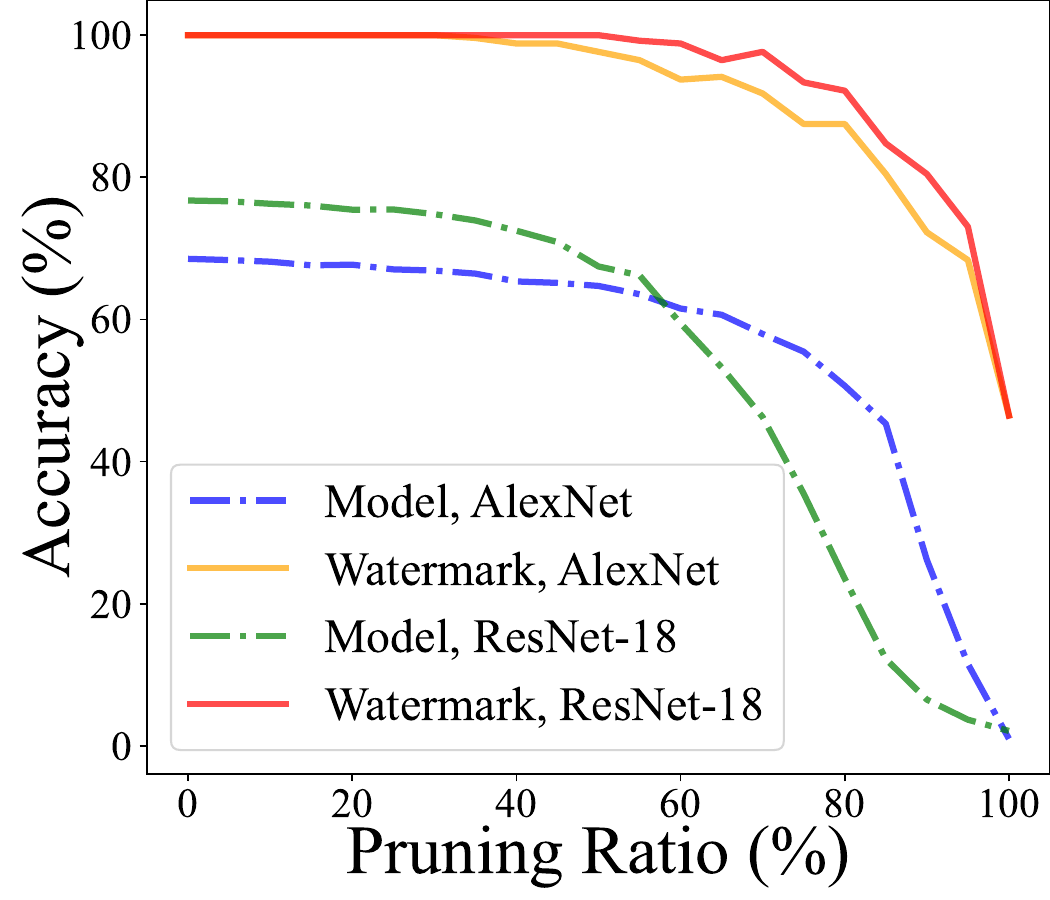}}
}
\caption{Comparison of resistance to pruning attacks at various pruning ratios on CIFAR-100 using AlexNet and ResNet-18.}
\label{appendix_fig:PruningAttackCIFAR-100}
\vspace{-1ex}
\end{figure}

\begin{figure}[htbp]
\centering
\subfloat[NeuralMark] 
{
{\includegraphics[width=0.22\textwidth]{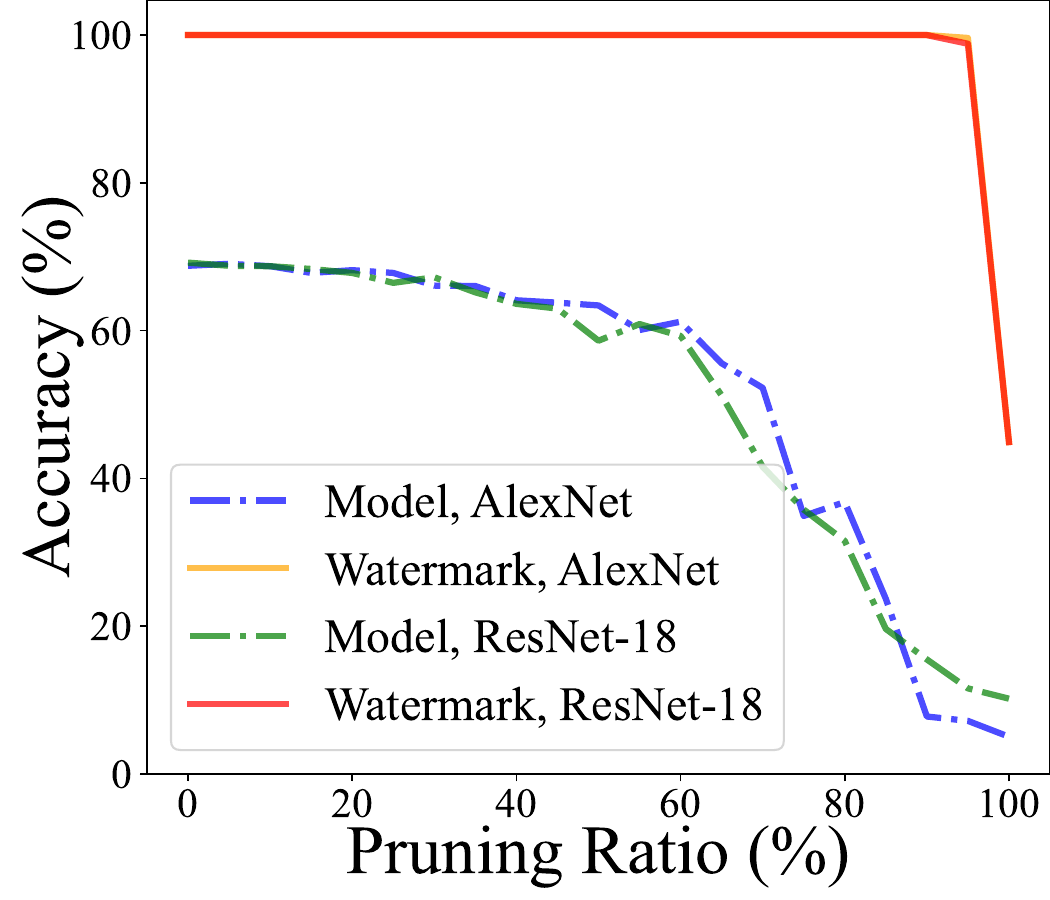}}
}
\hspace{-2.5mm}
\subfloat[VanillaMark] 
{
{\includegraphics[width=0.22\textwidth]{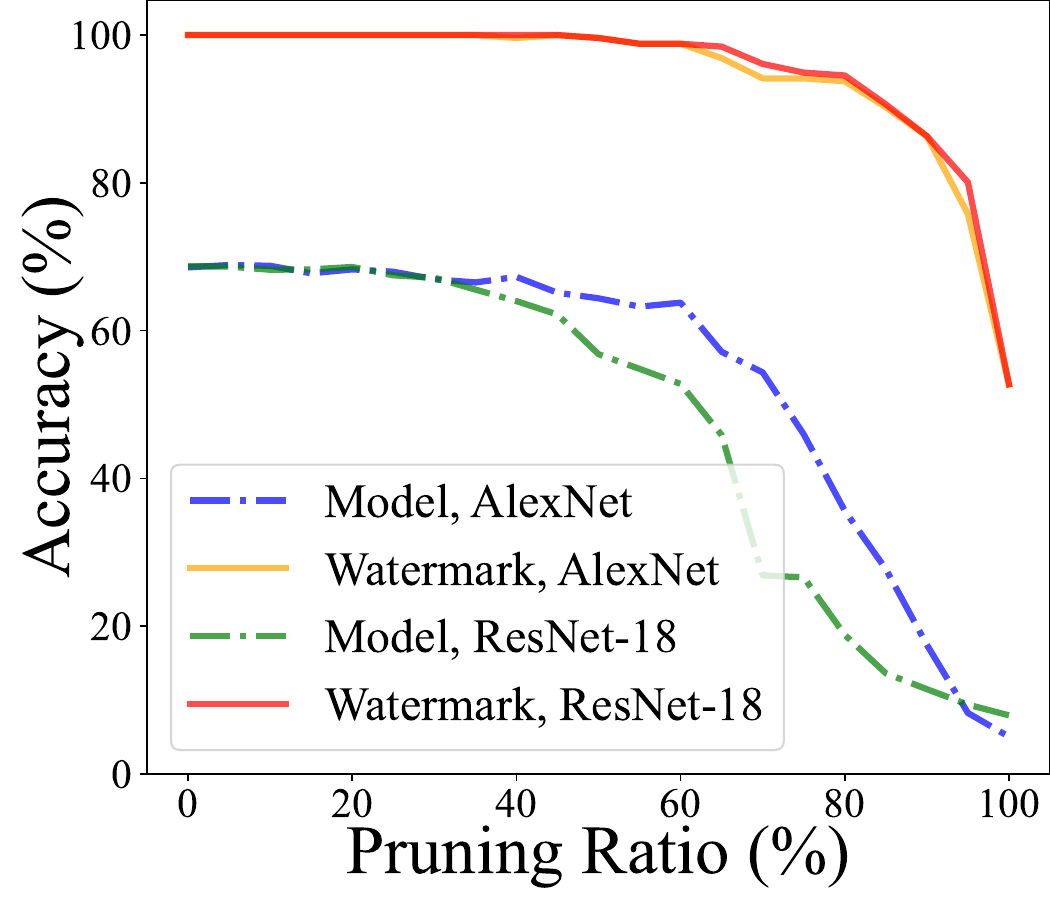}}
}
\hspace{-2.5mm}
\subfloat[GreedyMark]
{
{\includegraphics[width=0.22\textwidth]{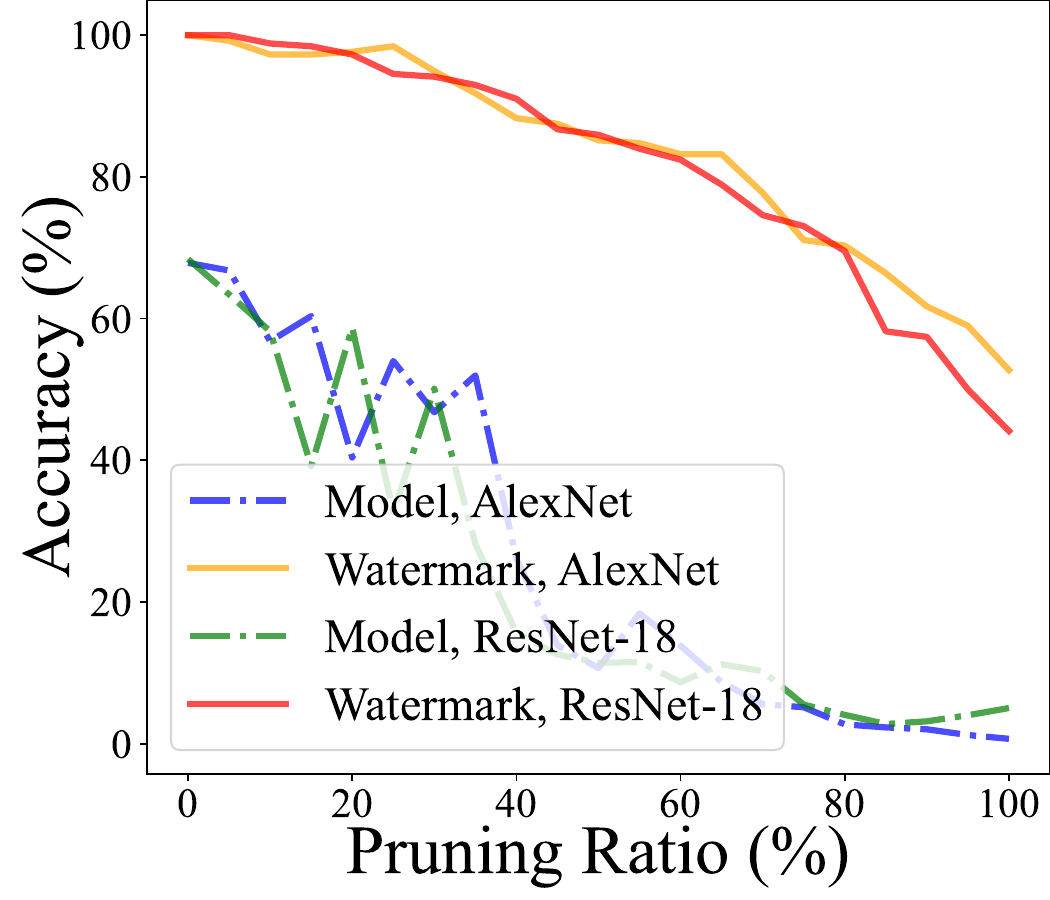}}
}
\hspace{-2.5mm}
\subfloat[VoteMark]
{
{\includegraphics[width=0.22\textwidth]{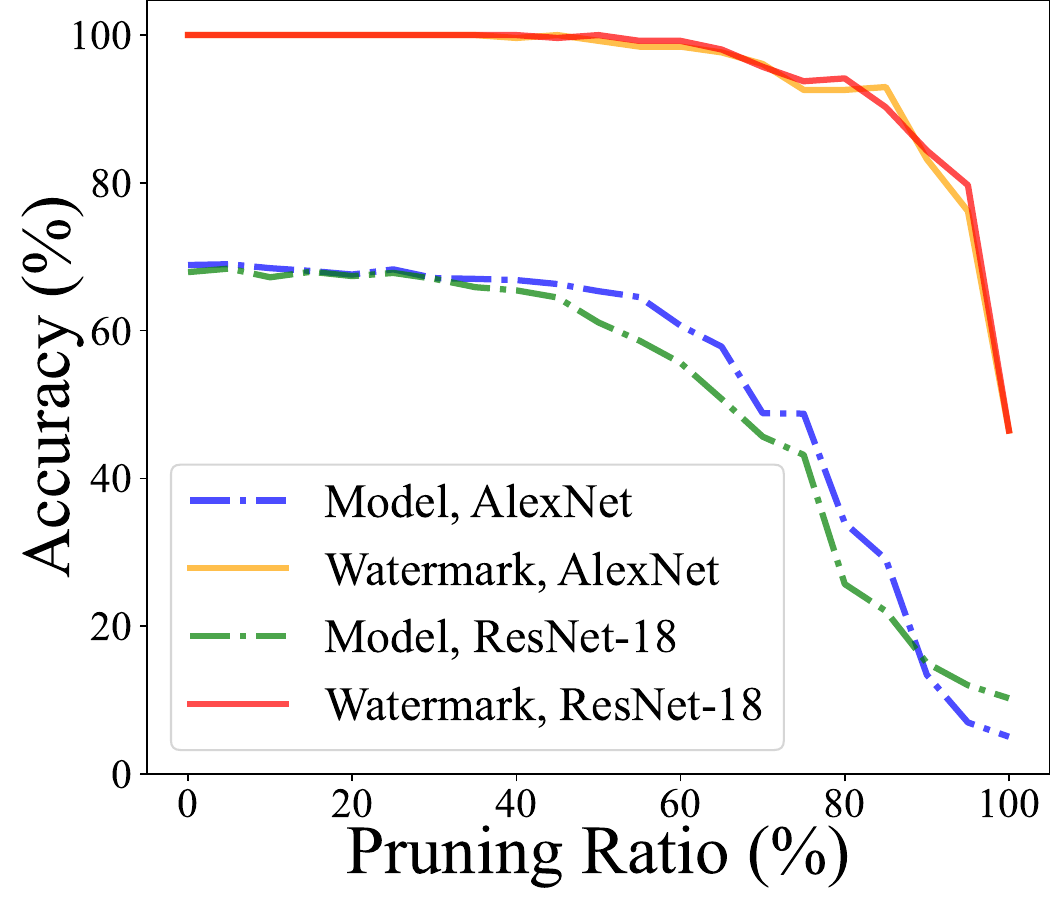}}
}
\caption{Comparison of resistance to pruning attacks at various pruning ratios on Caltech-101 using AlexNet and ResNet-18.}
\label{appendix_fig:PruningAttackCaltech-101}
\vspace{-1ex}
\end{figure}

\begin{figure}[htbp]
\centering
\subfloat[NeuralMark] 
{
{\includegraphics[width=0.22\textwidth]{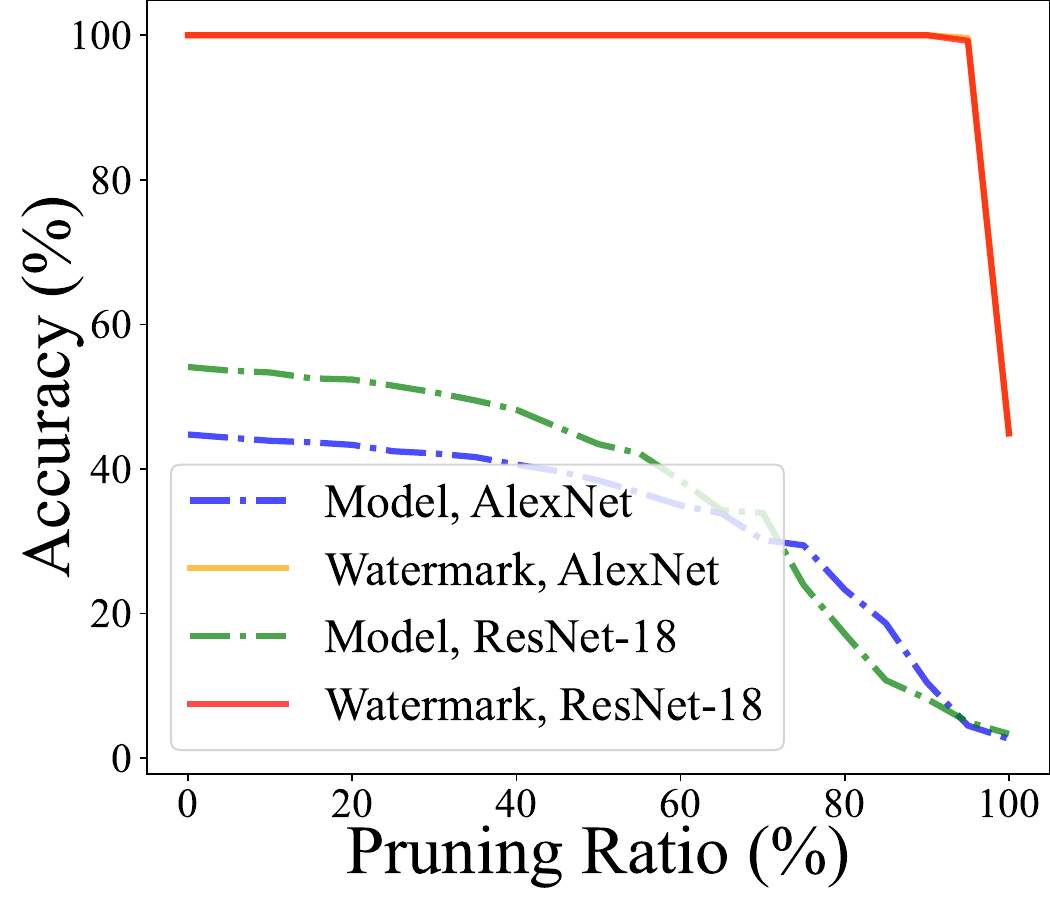}}
}
\hspace{-2.5mm}
\subfloat[VanillaMark] 
{
{\includegraphics[width=0.22\textwidth]{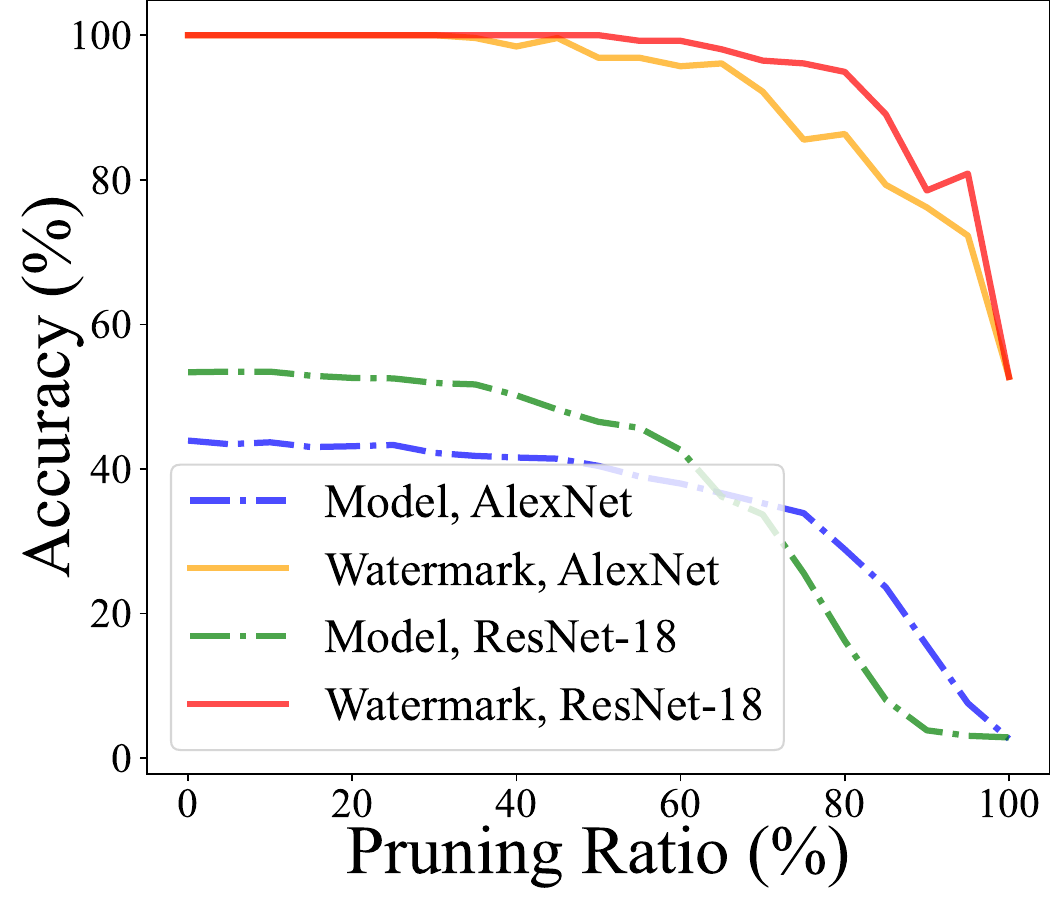}}
}
\hspace{-2.5mm}
\subfloat[GreedyMark]
{
{\includegraphics[width=0.22\textwidth]{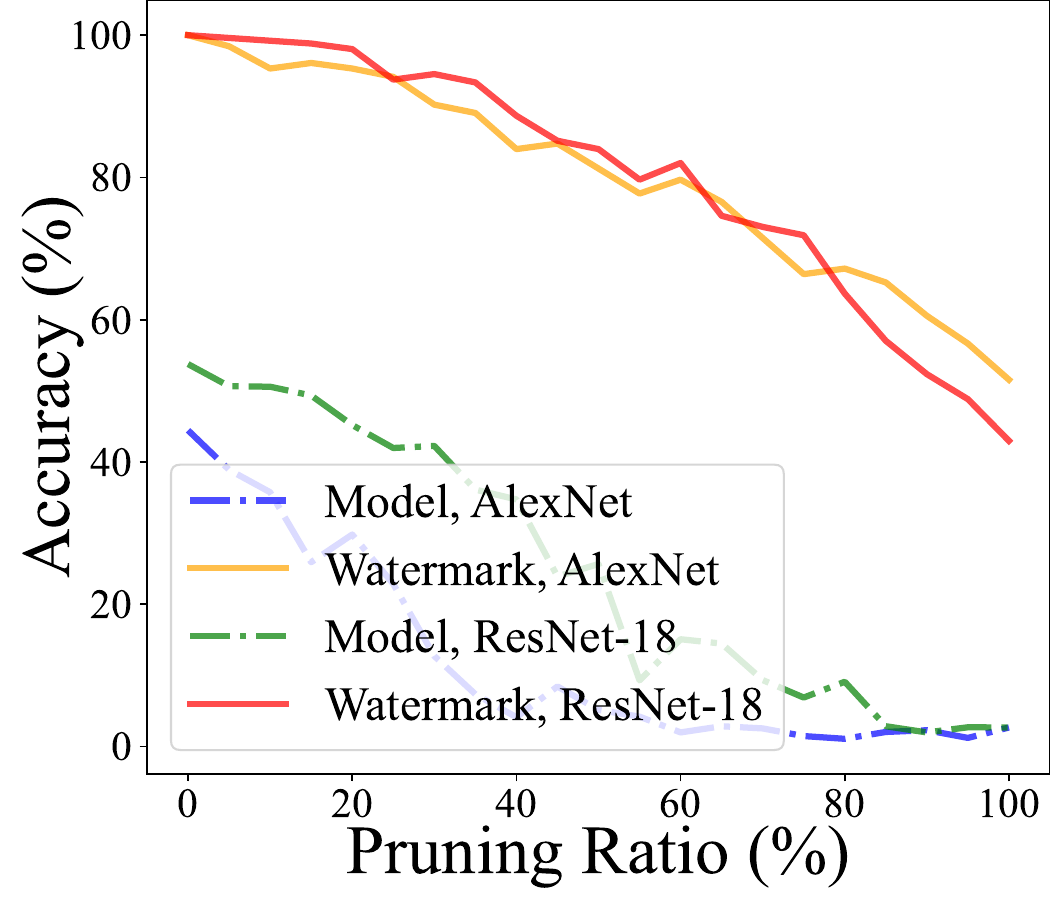}}
}
\hspace{-2.5mm}
\subfloat[VoteMark]
{
{\includegraphics[width=0.22\textwidth]{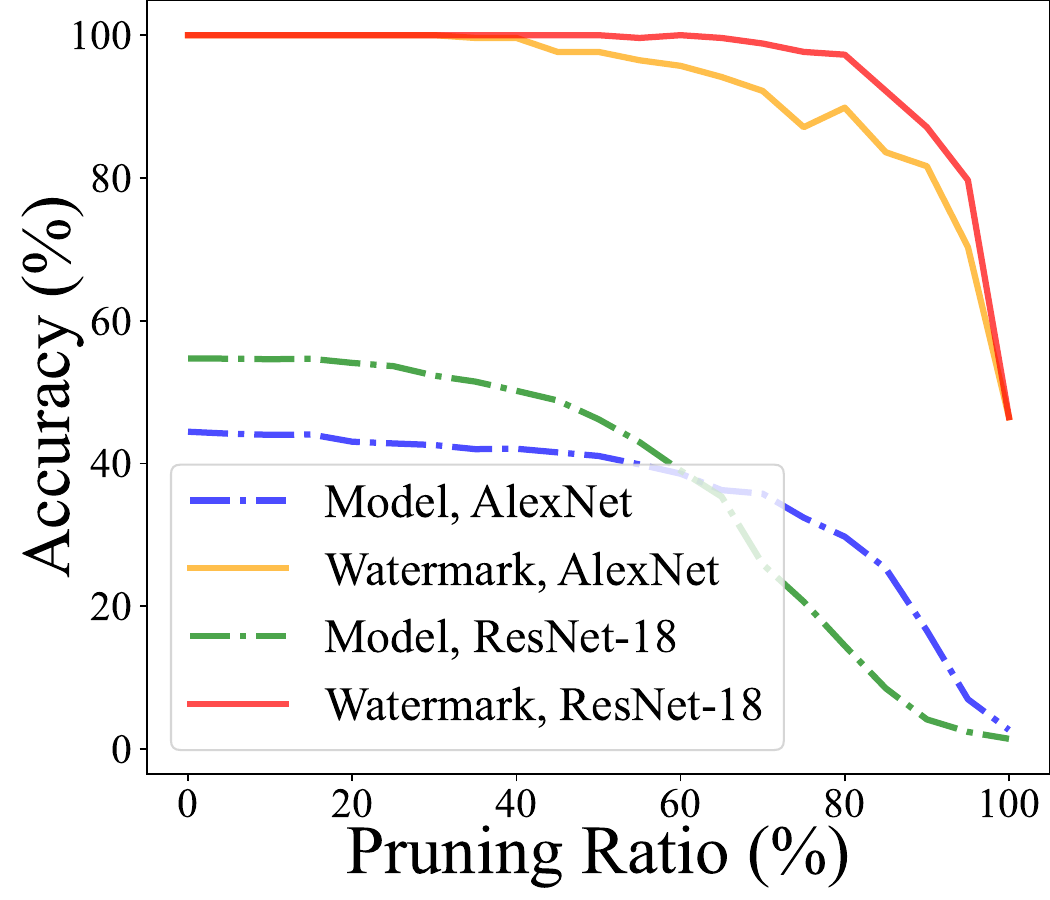}}
}
\caption{Comparison of resistance to pruning attacks at various pruning ratios on Caltech-256 using AlexNet and ResNet-18.}
\label{appendix_fig:PruningAttackCaltech-256}
\end{figure}

Figures~\ref{appendix_fig:PruningAttackCIFAR-10}-\ref{appendix_fig:PruningAttackCaltech-256} provide all the results from pruning attacks conducted on the CIFAR-10, CIFAR-100, Caltech-101, and Caltech-256 datasets, respectively.
As can be seen, as the pruning ratio increases, the performance of NeuralMark degrades while the detection rate remains nearly 100\%. This indicates NeuralMark's robustness against pruning attacks.
Those results collectively suggest NeuralMark exhibits superior robustness in resisting pruning attacks compared to other methods.


\subsection{F.4 Parameter Distribution}
\label{appendix:Distribution}

\begin{figure}[H]
\centering
\subfloat[AlexNet \label{appendix_fig:distributionAlexNet}] 
{
{\includegraphics[width=0.24\textwidth]{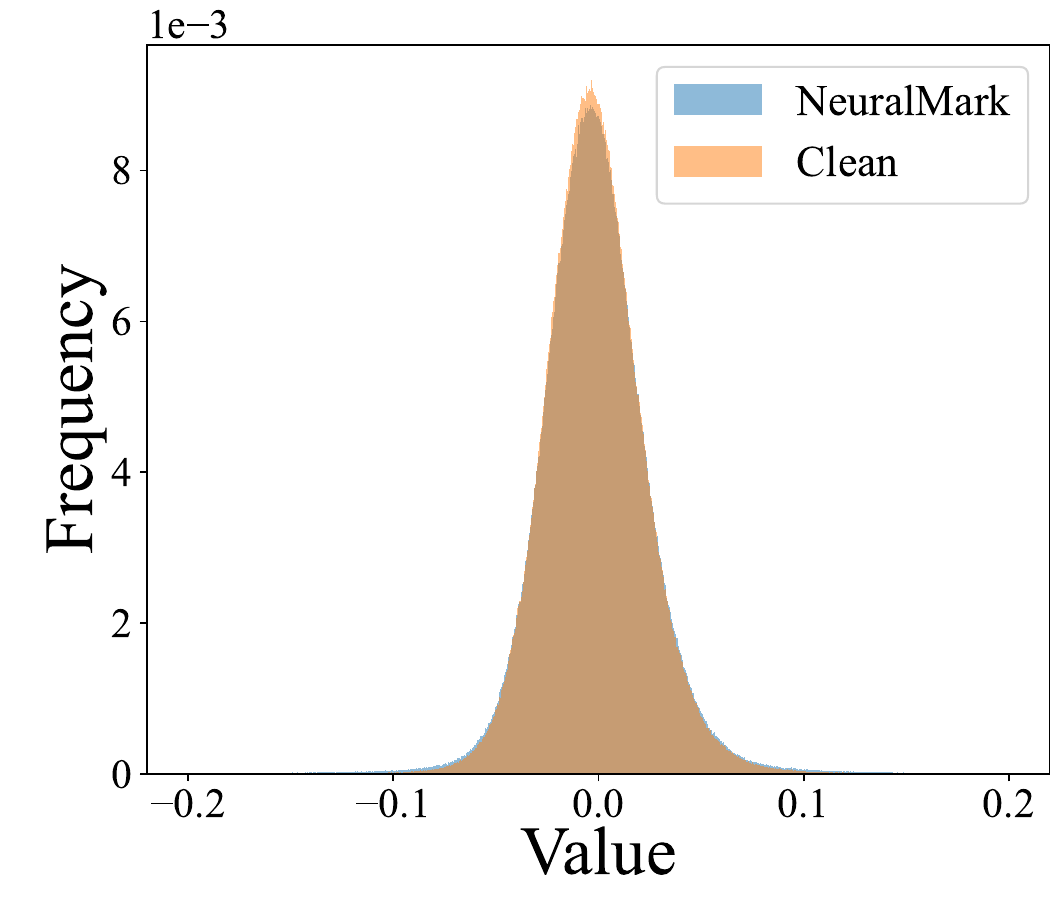}}
}
\hspace{-2.5mm}
\subfloat[ResNet-18 \label{appendix_fig:distributionResNet-18}] 
{
{\includegraphics[width=0.24\textwidth]{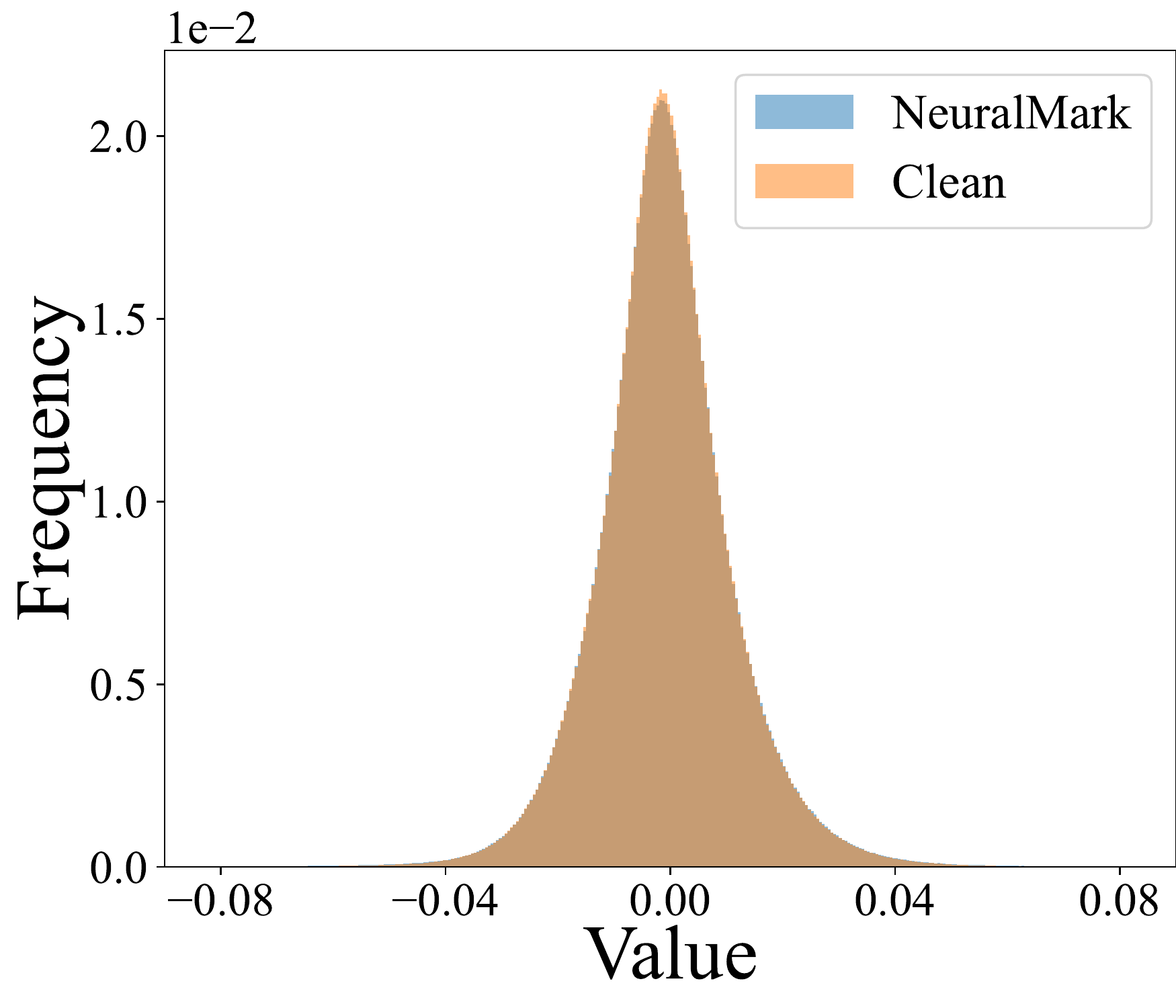}}
}
\hspace{-2.5mm}
\subfloat[ResNet-34 \label{appendix_fig:distributionResNet-34}] 
{
{\includegraphics[width=0.24\textwidth]{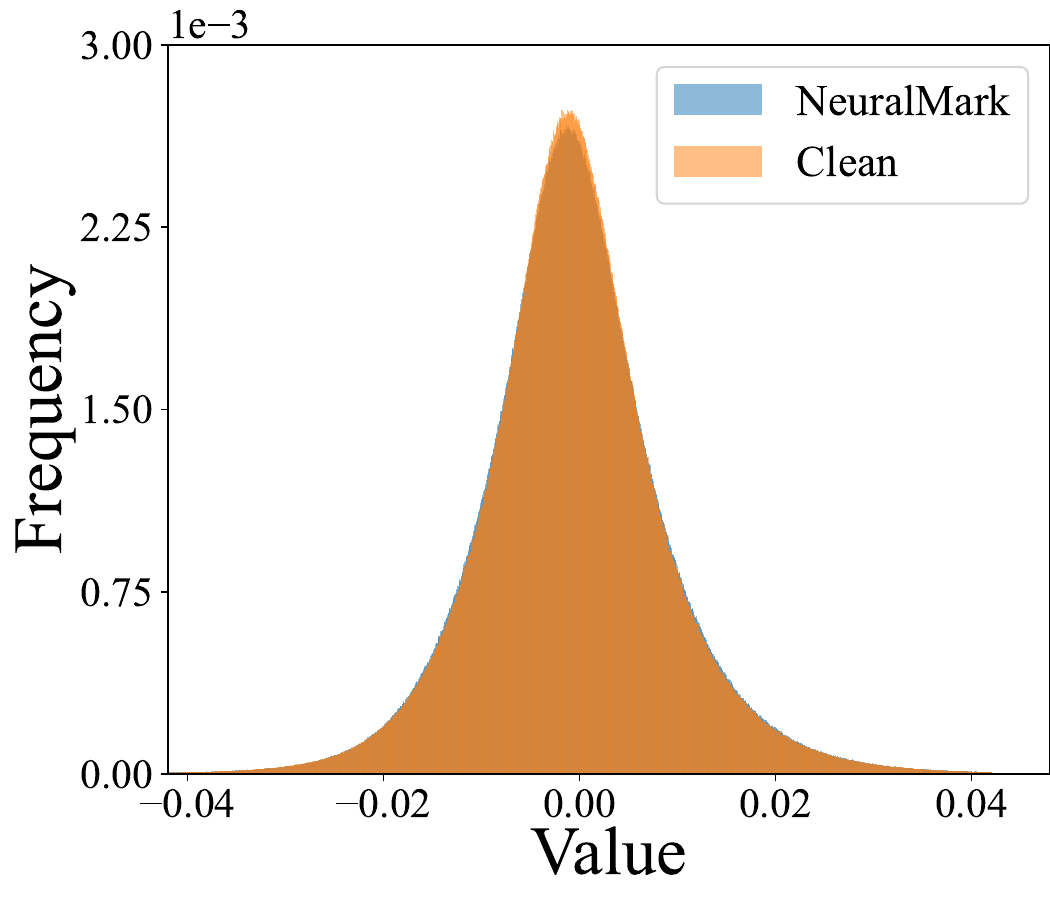}}
}
\hspace{-2.5mm}
\subfloat[ViT-B/16 \label{appendix_fig:distributionViT-B/16}]
{
{\includegraphics[width=0.24\textwidth]{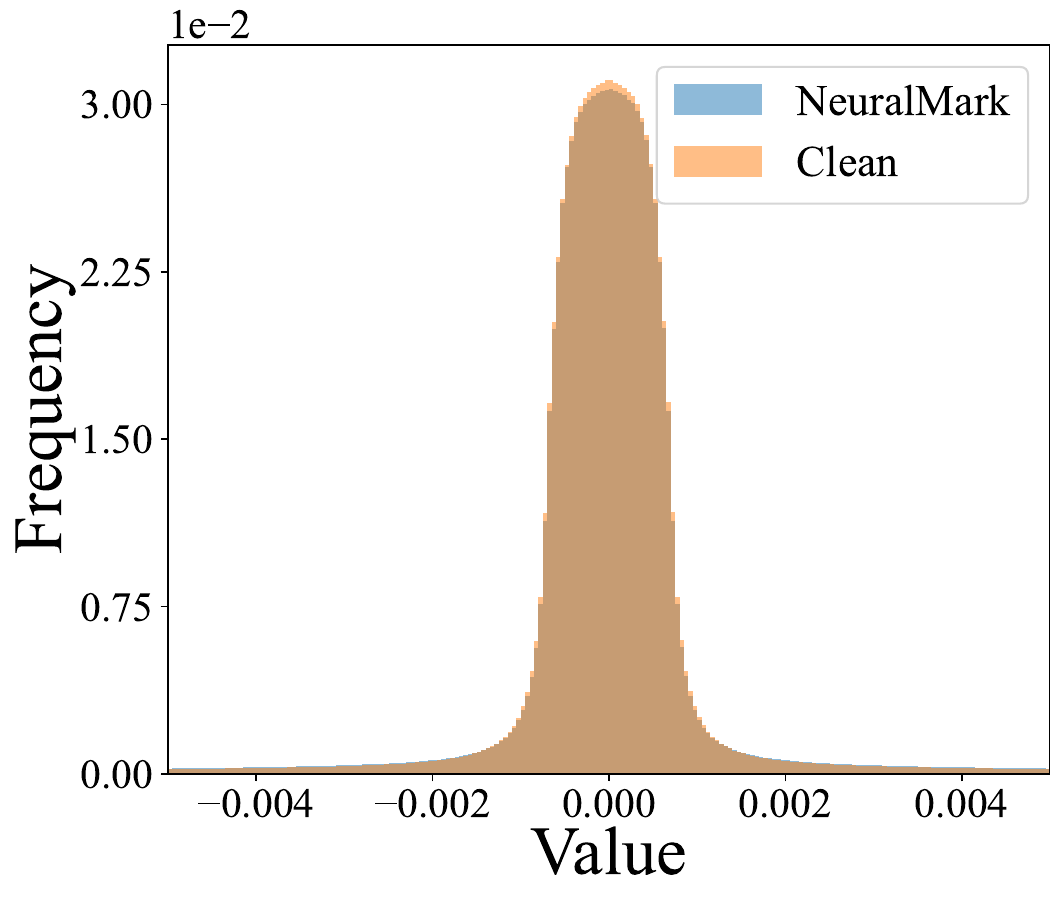}}
}
\hspace{-2.5mm}
\subfloat[VGG-16 \label{appendix_fig:distributionVGG-16}] 
{
{\includegraphics[width=0.24\textwidth]{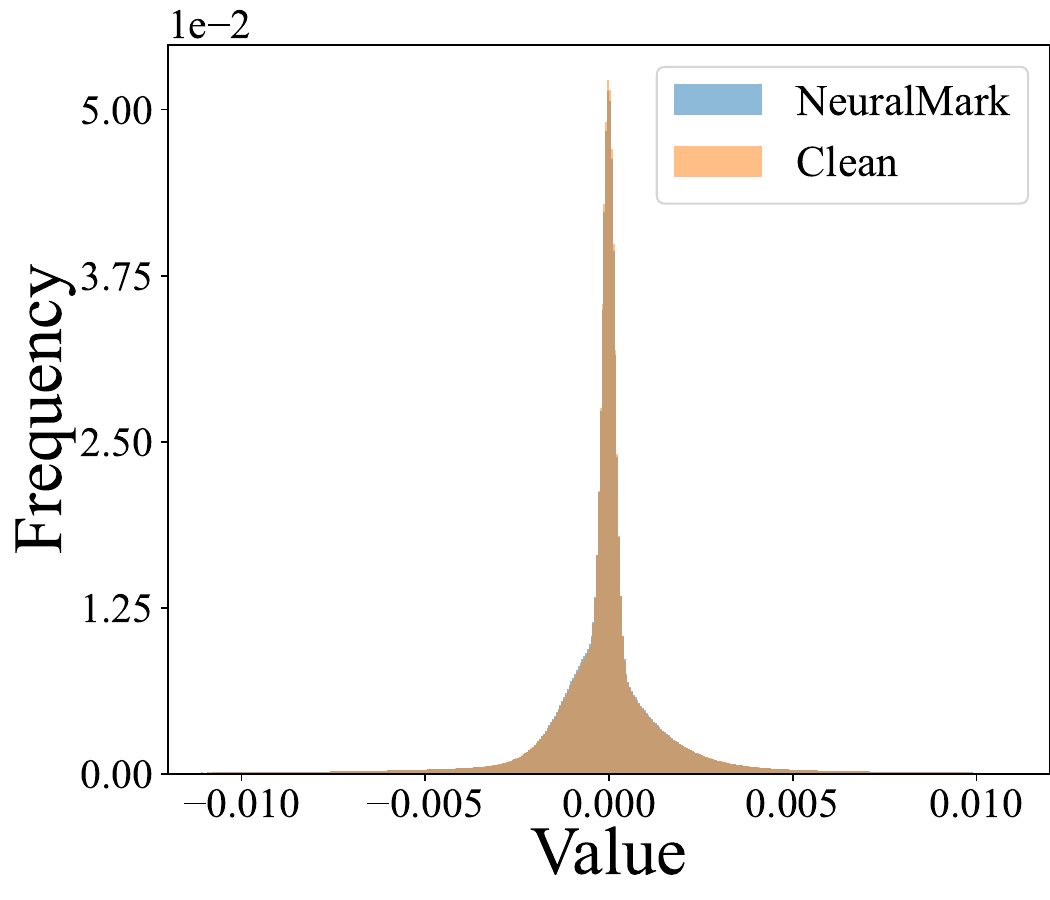}}
}
\hspace{-2.5mm}
\subfloat[MobileNet-V3-L \label{appendix_fig:distributionMobileNet-V3-L}] 
{
{\includegraphics[width=0.24\textwidth]{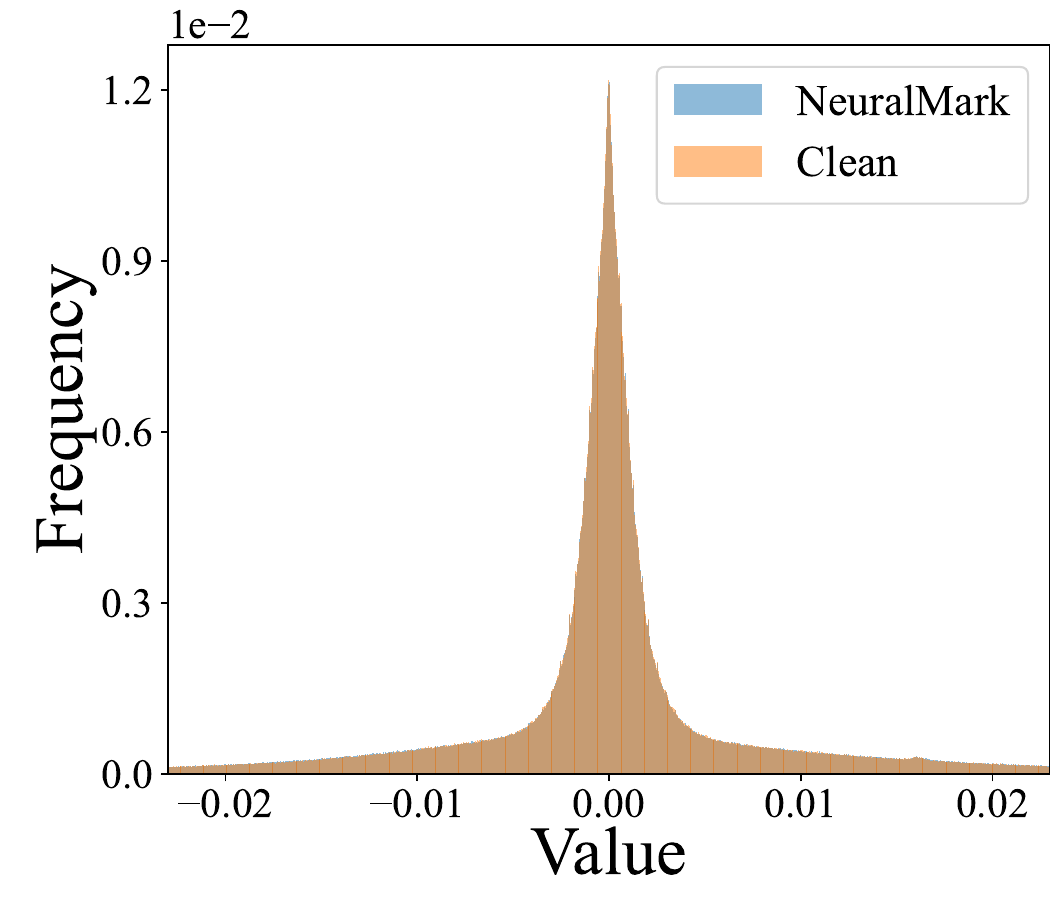}}
}
\hspace{-2.5mm}
\subfloat[GoogLeNet \label{appendix_fig:distributionGoogLeNet}]
{
{\includegraphics[width=0.24\textwidth]{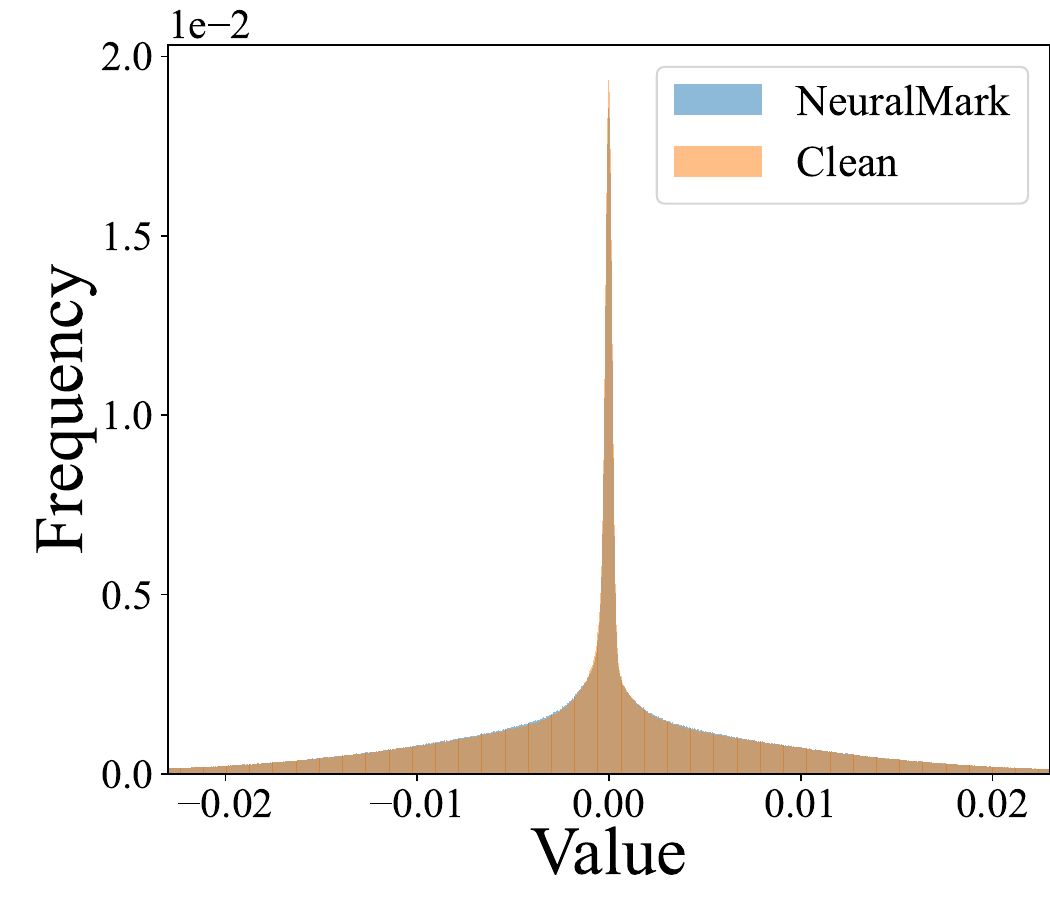}}
}
\hspace{-2.5mm}
\subfloat[Swin-V2-B \label{appendix_fig:distributionSwinV2}]
{
{\includegraphics[width=0.24\textwidth]{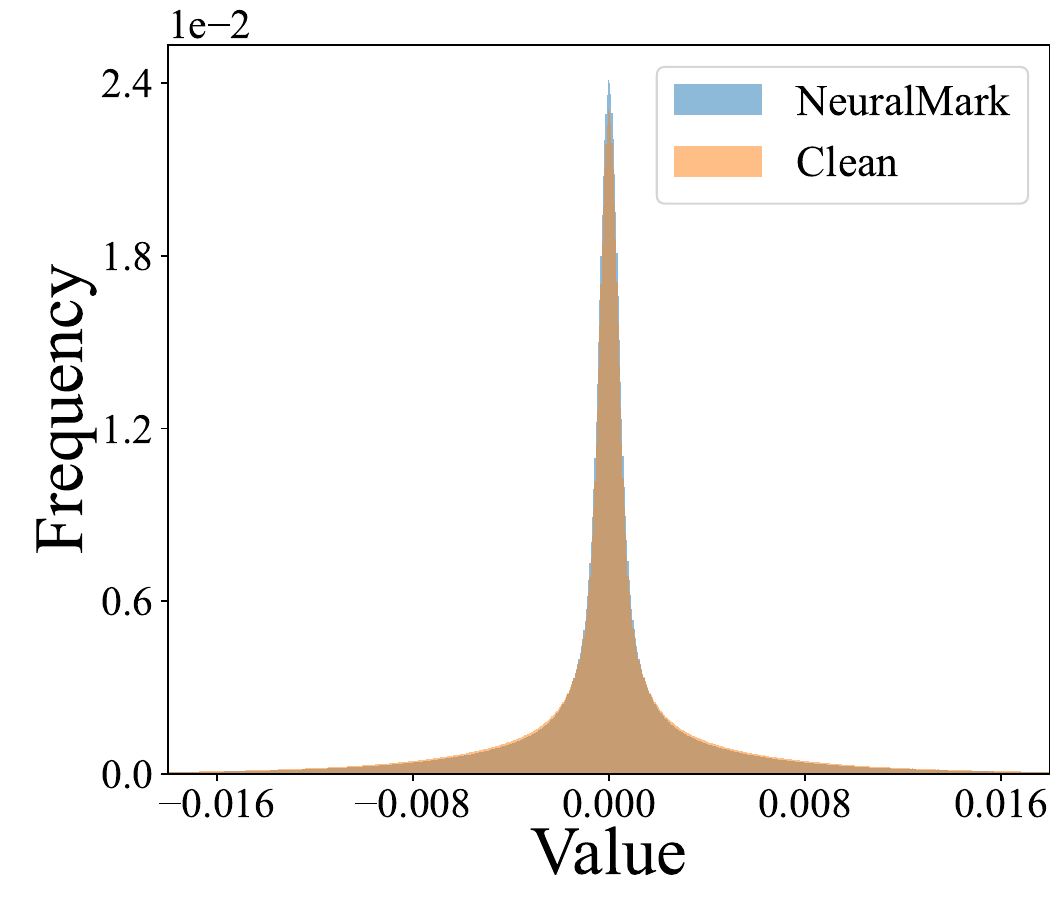}}
}
\caption{Comparison of parameter distributions on CIFAR-100 with distinct architectures.}
\label{appendix_fig:Distribution}
\end{figure}

\cref{appendix_fig:Distribution} provides additional parameter distributions for various architectures on the CIFAR-100 dataset. As can be seen, the parameter distributions of Clean and NeuralMark closely align in each architecture. Those results further demonstrate the secrecy of NeuralMark.

\subsection{F.5 Performance Convergence}
\label{appendix:Convergence}

\begin{figure}[H]
\centering
\subfloat[AlexNet \label{appendix_fig:ConvergenceAlextNet}] 
{
{\includegraphics[width=0.24\textwidth]{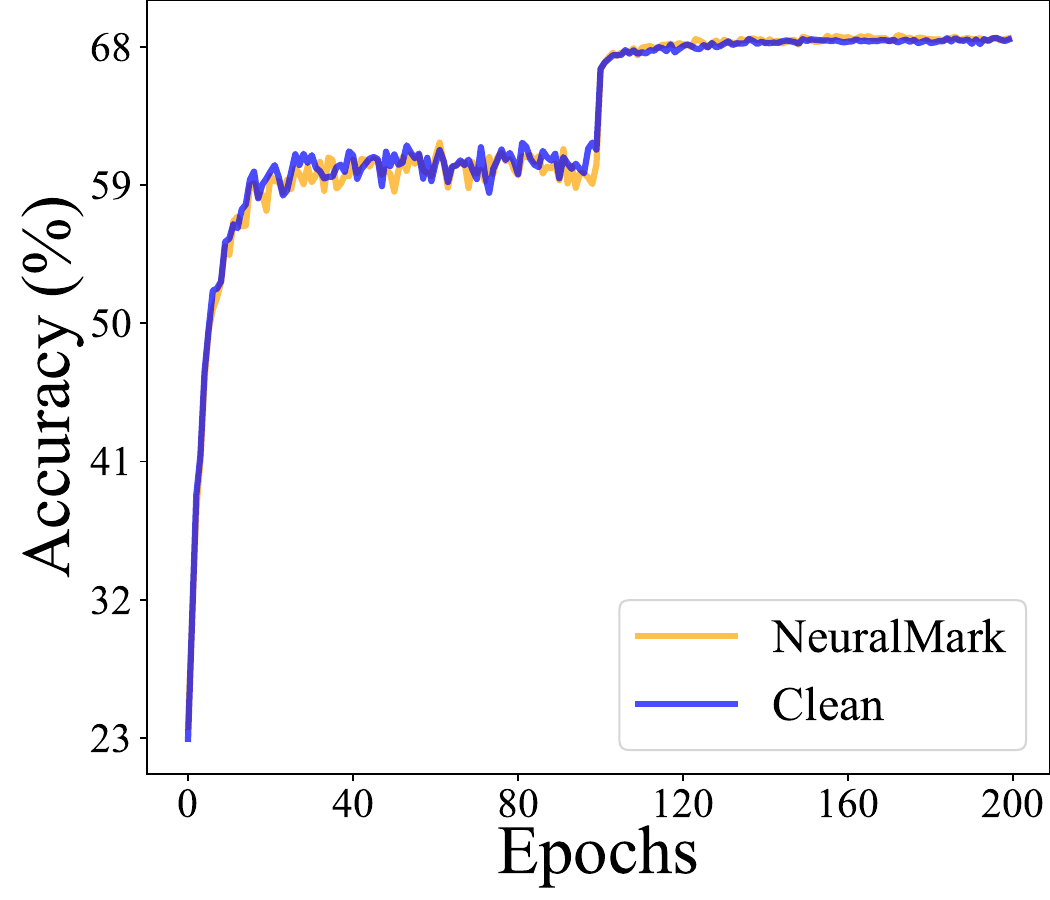}}
}
\hspace{-2.5mm}
\subfloat[ResNet-18 \label{appendix_fig:ConvergenceResNet-18}] 
{
{\includegraphics[width=0.24\textwidth]{./Figures/Analysis/Convergence/resnet.pdf}}
}
\hspace{-2.5mm}
\subfloat[ResNet-34 \label{appendix_fig:ConvergenceResNet-34}] 
{
{\includegraphics[width=0.24\textwidth]{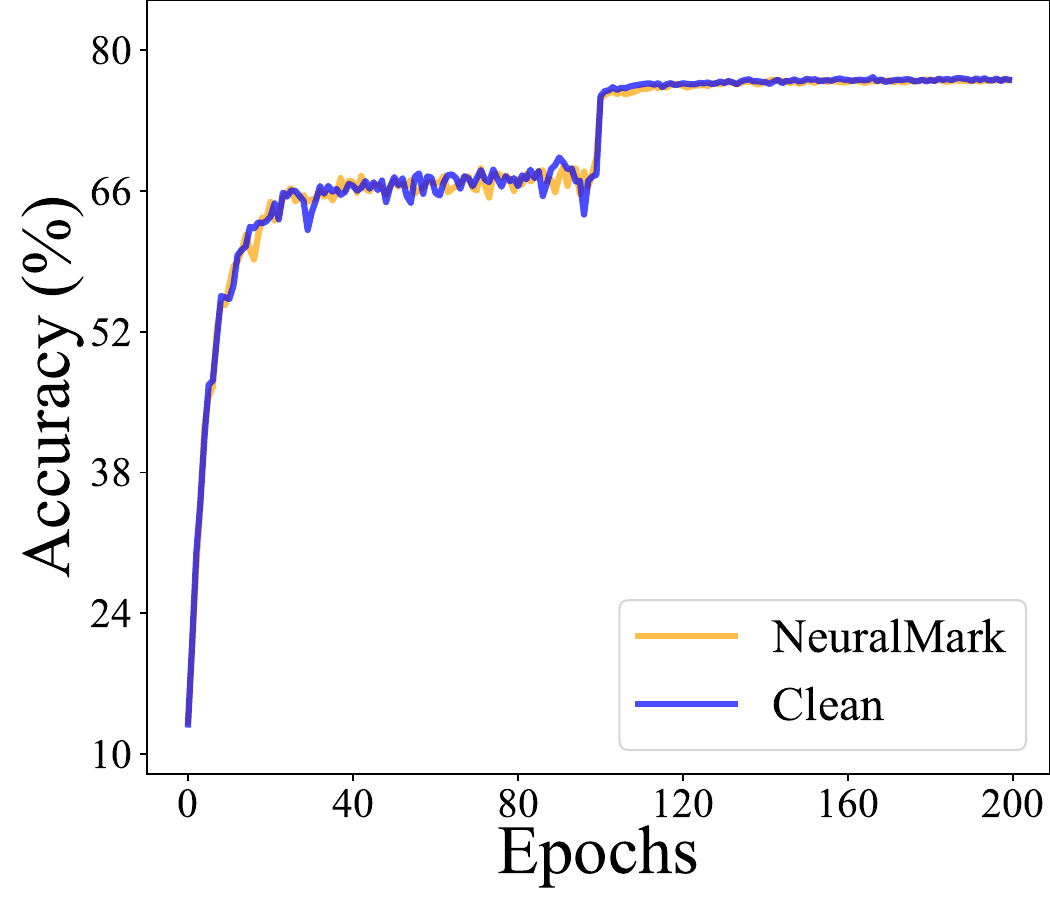}}
}
\hspace{-2.5mm}
\subfloat[ViT-B/16 \label{appendix_fig:ConvergenceViT-B/16}]
{
{\includegraphics[width=0.24\textwidth]{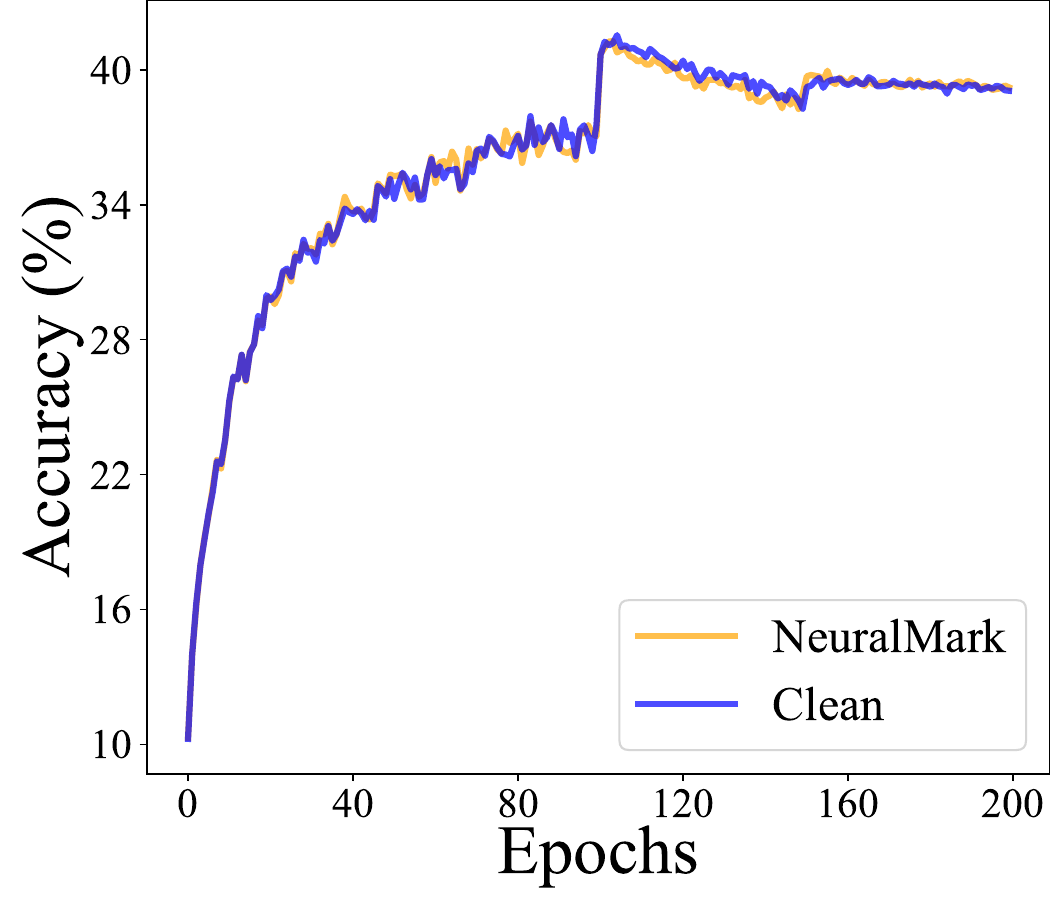}}
}
\hspace{-2.5mm}
\subfloat[VGG-16 \label{appendix_fig:ConvergenceVGG-16}] 
{
{\includegraphics[width=0.24\textwidth]{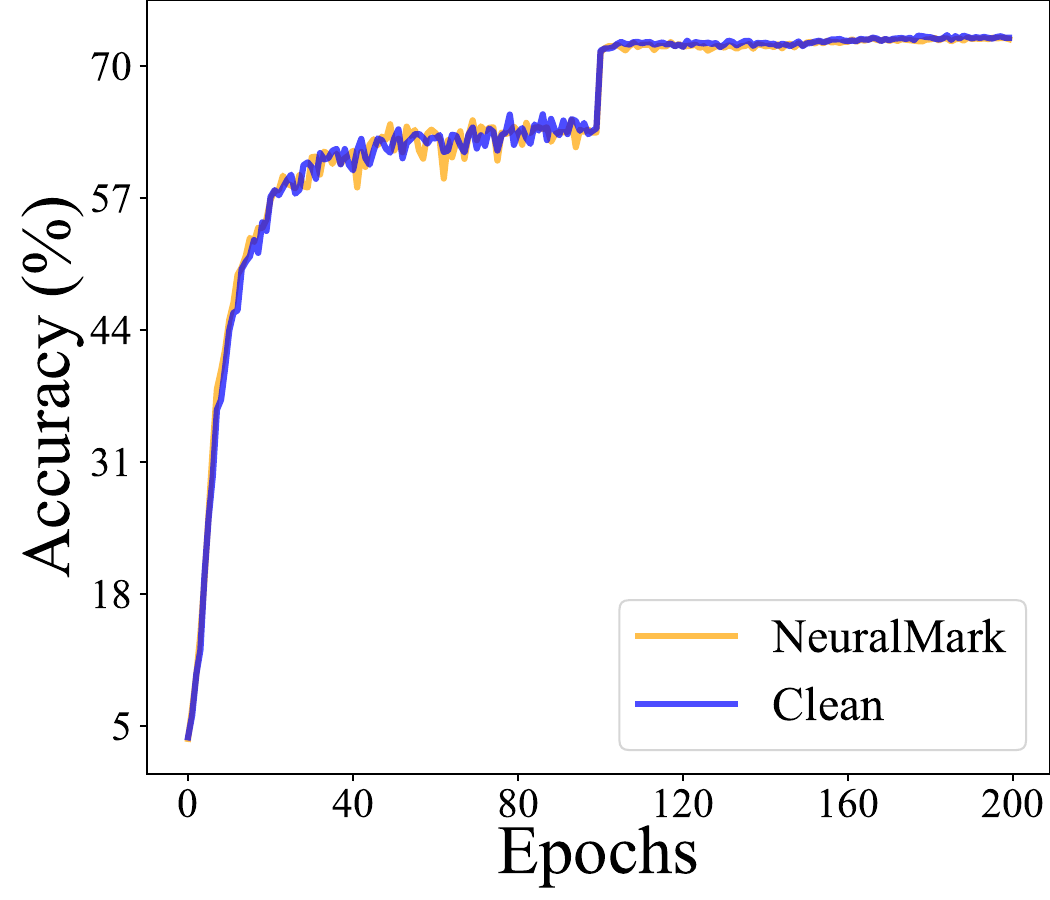}}
}
\hspace{-2.5mm}
\subfloat[MobileNet-V3-L \label{appendix_fig:ConvergenceMobileNet-V3-L}] 
{
{\includegraphics[width=0.24\textwidth]{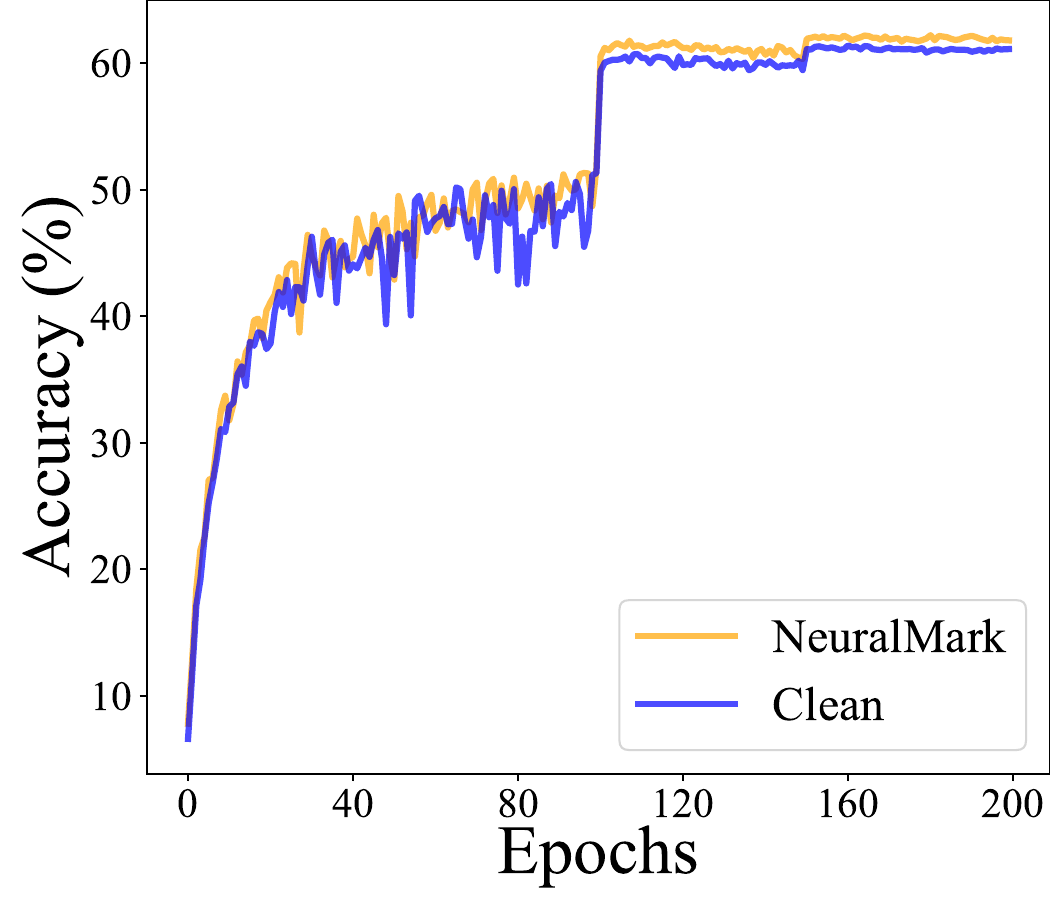}}
}
\hspace{-2.5mm}
\subfloat[GoogLeNet \label{appendix_fig:ConvergenceGoogLeNet}]
{
{\includegraphics[width=0.24\textwidth]{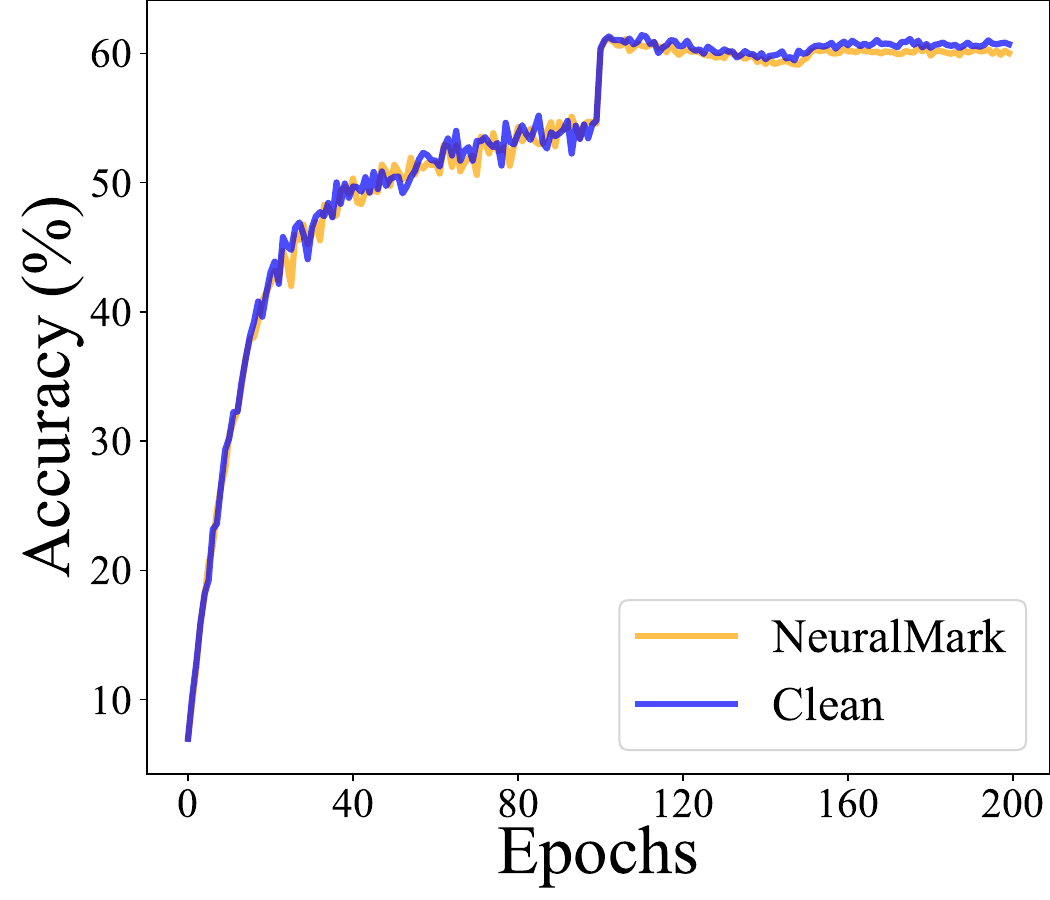}}
}
\hspace{-2.5mm}
\subfloat[Swin-V2-B \label{fig:ConvergenceSwin-V2-B}]
{
{\includegraphics[width=0.24\textwidth]{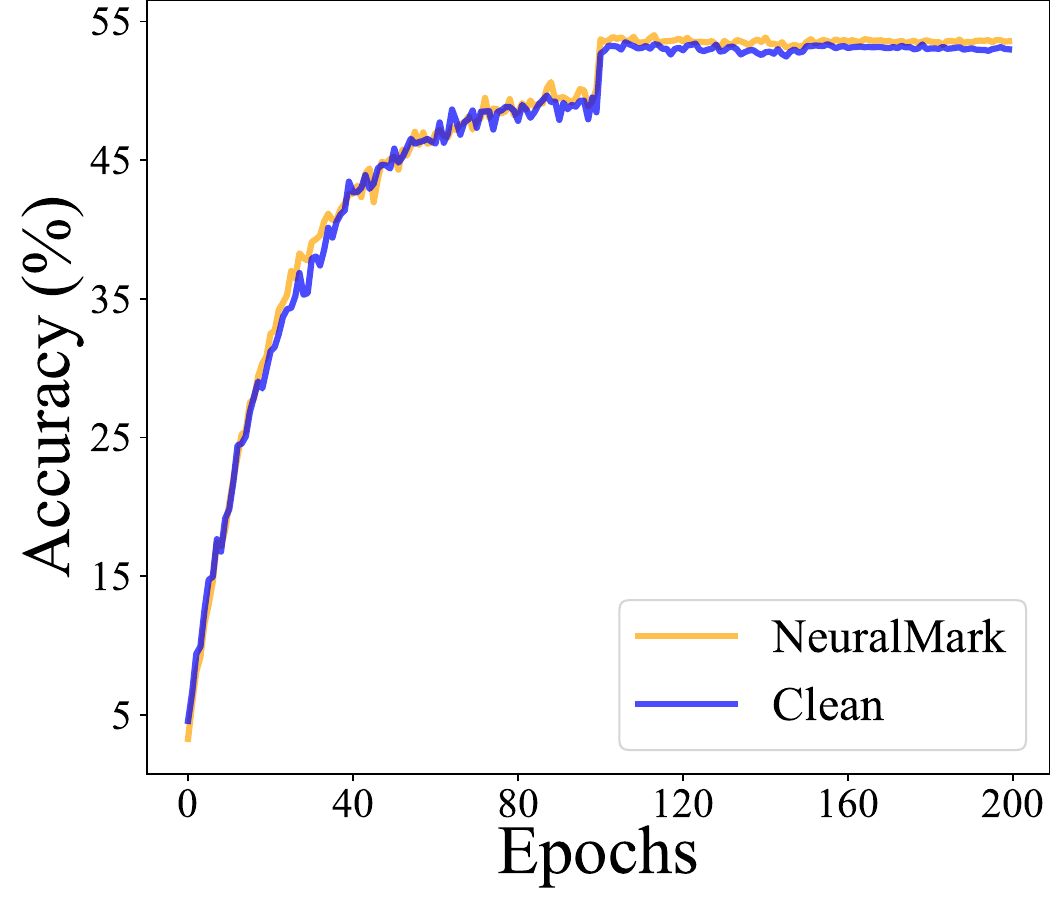}}
}
\caption{Comparison of model performance convergence across distinct architectures on CIFAR-100.}
\label{appendix_fig:Convergence}
\end{figure}

\cref{appendix_fig:Convergence} presents additional performance convergence plots for various architectures on the CIFAR-100 dataset. Across all architectures, the performance curves of Clean and NeuralMark exhibit similar trends and are closely aligned, further confirming that NeuralMark does not negatively affect performance convergence.

\subsection{F.6 Watermark Embedding Layers}

To investigate the impact of watermark embedding layers on the model performance, we randomly choose four individual layers and all layers from ResNet-18 for watermark embedding. \cref{appendix_tab:Layers} presents the results on CIFAR-100, showing that embedding different layers or all layers does not significantly affect the model performance.

\begin{table*}[htbp]
  \centering
   \caption{Comparison of classification accuracy (\%) on different watermarking layers on CIFAR-100 using ResNet-18. Here, Layers 1-4 denote randomly chosen layers, while All Layers refers to all layers. Watermark detection rates are omitted as they all reach 100\%.}
  \resizebox{0.65\columnwidth}{!}{
    \begin{tabular}{c|ccccc}
    \toprule
     Watermarking Layer & Layer 1 & Layer 2 & Layer 3 & Layer 4 & All Layers \\
    \midrule
    Accuracy & 76.51  & 76.68  & 76.30 & 76.73  & 75.86  \\
    \bottomrule
    \end{tabular}%
    }
  \label{appendix_tab:Layers}%
  \vspace{-2ex}
\end{table*}%

\subsection{F.7 Watermark Length} 

To evaluate the influence of watermark length on the model performance, we set watermark lengths to 64, 128, 256, 512, 1024, and 2048, respectively. \cref{appendix_tab:length} lists the results on the CIFAR-100 dataset using ResNet-18, indicating that NeuralMark can achieve a $100\%$ detection rate with various watermark lengths while preserving nearly lossless model performance.

\begin{table}[H]
  \centering
  \setlength{\tabcolsep}{2.8mm}
  \caption{Comparison of classification accuracy (\%) for distinct watermark lengths on CIFAR-100 using ResNet-18. Watermark detection rates are omitted as they all reach $100\%$.}
  \resizebox{0.65\columnwidth}{!}{
  \begin{tabular}{c|cccccc}
    \toprule
    Watermark Length & 64    & 128   & 256   & 512   & 1024  & 2048 \\
    \midrule
    Accuracy & 75.84  & 75.90  & 76.46  & 76.18  & 76.51  & 76.27  \\
    \bottomrule
    \end{tabular}%
    }
  \label{appendix_tab:length}%
  \vspace{-1ex}
\end{table}%

\subsection{F.8 Training Efficiency} 

\cref{appendix_tab:efficieny} lists the average time cost (in seconds) per training epoch over five epochs on the CIFAR-100 dataset using ResNet-18. 
NeuralMark's running time is comparable to that of Clean and VanillaMark, highlighting the efficiency of NeuralMark. Additionally, NeuralMark outperforms GreedyMark in terms of speed, as GreedyMark relies on costly sorting operations for parameter selection. Moreover, NeuralMark demonstrates significantly faster running times compared to VoteMark, as it avoids the multiple rounds of watermark embedding loss calculations required by VoteMark. Those results highlight the superior efficiency of NeuralMark.

\begin{table}[H]
  \centering
  \vspace{-1ex}
  \caption{Comparison of average time cost (in seconds) on CIFAR-100 using ResNet-18. Here, $R$ denotes the number of filtering rounds.}
  \resizebox{\textwidth}{!}{
    \begin{tabular}{c|cccccccc}
    \toprule
    Method & Clean & \parbox{2cm}{NeuralMark \\ \centering ($R = 1$)}  & \parbox{2cm}{NeuralMark \\ \centering ($R = 2$)} & \parbox{2cm}{NeuralMark \\ \centering ($R = 3$)} & \parbox{2cm}{NeuralMark \\ \centering ($R = 4$)} & VanillaMark & GreedyMark & VoteMark \\
    \midrule
    Time (s) & 23.60  & 24.49  & 24.94  & 25.01  & 25.19  & 24.34  & 47.43 & 35.17 \\
    \bottomrule
    \end{tabular}%
    }
  \label{appendix_tab:efficieny}%
\end{table}%

\section{G. Further Analysis on Filtering Rounds}
\label{appendix:Filtering}

To assess the influence of the number of filtering rounds on NeuralMark's effectiveness and robustness in resisting various attacks, we conduct additional experiments using 6 and 8 filters, compared to NeuralMark’s default setting of
4 filters. We omit forging attacks, since the hashed watermark filter is intrinsically resistant to such attacks.

\subsection{G.1 Fidelity Evaluation}

\cref{appendix_tab:fidelityMultiFiltering} presents the impact of watermark embedding on the model performance across distinct filtering rounds. The results demonstrate that NeuralMark, even with varying filtering rounds, has a minimal effect on the model performance while successfully embedding watermarks. 
\begin{table*}[htbp]
  \centering
  \vspace{-1ex}
    \caption{Comparison of classification accuracy (\%) with various distinct filter rounds on CIFAR-10 and CIFAR-100 using ResNet-18. Watermark detection rates are omitted as they all reach 100\%.}
    \resizebox{0.4\columnwidth}{!}{
    \begin{tabular}{c|ccc}
    \toprule
    Dataset &   4 Filters &   6 Filters &  8 Filters \\
    \midrule
    CIFAR-10 & 94.79 & 94.74  & 94.88  \\
    CIFAR-100 & 76.74 & 75.59  & 76.16  \\
    \bottomrule
    \end{tabular}%
    }
  \label{appendix_tab:fidelityMultiFiltering}%
  \vspace{-1ex}
\end{table*}%

\subsection{G.2 Robustness Evaluation}

\subsubsection{Fine-tuning Attacks on All Model Parameters} 

\cref{appendix_tab:fine-tuningMultiFiltering} reports the results of fine-tuning attacks across distinct filtering rounds. The attacks are performed by updating all model parameters with a learning rate of 0.001. As shown, NeuralMark maintains a watermark detection rate of 100\% across all filtering rounds, with negligible impact on model performance.

\begin{table}[htbp]
  \centering
  \vspace{-1ex}
  \caption{Comparison of resistance to fine-tuning attacks with distinct filter rounds using ResNet-18. Watermark detection rates are omitted as they all reach 100\%.}
  \resizebox{0.6\columnwidth}{!}{
    \begin{tabular}{c|cccc}
    \toprule
    Fine-tuning & Clean &   4 Filters &   6 Filters &  8 Filters \\
    \midrule
    CIFAR-100 to CIFAR-10 & 93.21 & 93.74 & 93.01 & 93.55 \\
    CIFAR-10 to CIFAR-100 & 72.17 & 71.67 & 72.68 & 72.27 \\
    \bottomrule
    \end{tabular}%
    }
  \label{appendix_tab:fine-tuningMultiFiltering}%
  \vspace{-1ex}
\end{table}%

\subsubsection{Overwriting Attacks}

\cref{appendix_tab:overwritingMultiFiltering} lists the results of overwriting attacks across distinct filtering rounds. From the results, we find that when the number of filtering rounds is set to 6, NeuralMark exhibits superior robustness compared to 4 and 8 filter rounds. Specifically, at $\eta = 0.01$, the original watermark detection rates for 4, 6, and 8 filter rounds on the CIFAR-100 to CIFAR-10 task are 92.18\%, 94.92\%, and 89.84\%, respectively. Those results indicate that increasing the number of filtering rounds can enhance robustness against overwriting attacks to a certain extent. However, when the number of filtering rounds exceeds a certain threshold, the robustness may be slightly compromised due to the reduction in the number of parameters.

\begin{table}[htbp]
\vspace{-1ex}
  \centering
  \caption{Comparison of resistance to overwriting attacks at various trade-off hyper-parameters ($\lambda$) and learning rates ($\eta$) with distinct filtering rounds using ResNet-18. Values (\%) \textbf{inside} and outside the bracket are \textbf{watermark detection rate} and classification accuracy, respectively.}
  \resizebox{\textwidth}{!}{
    \begin{tabular}{c|c|ccc|c|ccc}
    \toprule
    Overwriting & $\lambda$ &   4 Filters &   6 Filters &  8 Filters & $\eta$ &   4 Filters &   6 Filters &  8 Filters \\
    \midrule
    \multirow{5}[2]{*}{\parbox{2cm}{CIFAR-100 \\ \centering to \\CIFAR-10}} & 1     & 93.65 (\textbf{100}) & 93.13(\textbf{100}) & 93.40(\textbf{100}) & 0.001 & 93.65 (\textbf{100}) & 93.13(\textbf{100}) & 93.40(\textbf{100}) \\
          & 10    & 93.44 (\textbf{100}) & 93.06(\textbf{100}) & 93.41(\textbf{100}) & 0.005 & 91.76 (\textbf{99.60}) & 92.10(\textbf{100}) & 91.62(\textbf{100}) \\
          & 50    & 93.46 (\textbf{100}) & 93.06(\textbf{100}) & 93.54(\textbf{100}) & 0.01  & 91.58 (\textbf{92.18}) & 91.64(\textbf{94.92}) & 90.48(\textbf{89.84}) \\
          & 100   & 93.53 (\textbf{100}) & 92.88(\textbf{100}) & 92.99(\textbf{100}) & 0.1   & 75.2 (\textbf{50.78}) & 75.84(\textbf{58.2}) & 74.54(\textbf{51.56}) \\
          & 1000  & 93.09 (\textbf{100}) & 93.03(\textbf{100}) & 93.39(\textbf{100}) & 1     & 10.00 (\textbf{44.53}) & 10.00(\textbf{47.26}) & 10.00(\textbf{50.39}) \\
    \midrule
    \multirow{5}[2]{*}{\parbox{2cm}{CIFAR-10 \\ \centering to \\CIFAR-100}} & 1     & 71.78 (\textbf{100}) & 71.69(\textbf{100}) & 72.63(\textbf{100}) & 0.001 & 71.78 (\textbf{100}) & 71.69(\textbf{100}) & 72.63(\textbf{100}) \\
          & 10    & 72.6 (\textbf{100}) & 72.06(\textbf{100}) & 72.81(\textbf{100}) & 0.005 & 71.04 (\textbf{99.60}) & 70.65(\textbf{100}) & 71.46(\textbf{100}) \\
          & 50    & 72.73 (\textbf{100}) & 71.85(\textbf{100}) & 72.85(\textbf{100}) & 0.01  & 69.14 (\textbf{96.48}) & 69.47(\textbf{97.26}) & 67.88(\textbf{95.70}) \\
          & 100   & 71.49 (\textbf{100}) & 71.88(\textbf{100}) & 72.00(\textbf{100}) & 0.1   & 51.88 (\textbf{60.54}) & 55.18(\textbf{62.10}) & 50.36(\textbf{55.07}) \\
          & 1000  & 71.81 (\textbf{100}) & 72.22(\textbf{100}) & 72.39(\textbf{100}) & 1     & 1.00 (\textbf{44.53}) & 1.00(\textbf{47.26}) & 1.00(\textbf{50.39}) \\
    \bottomrule
    \end{tabular}%
    }
  \label{appendix_tab:overwritingMultiFiltering}%
  \vspace{-1ex}
\end{table}%

\subsubsection{Pruning Attacks} 

\begin{figure}[H]
\vspace{-1px}
\centering
\subfloat[CIFAR-10] 
{
{\includegraphics[width=0.26\columnwidth]{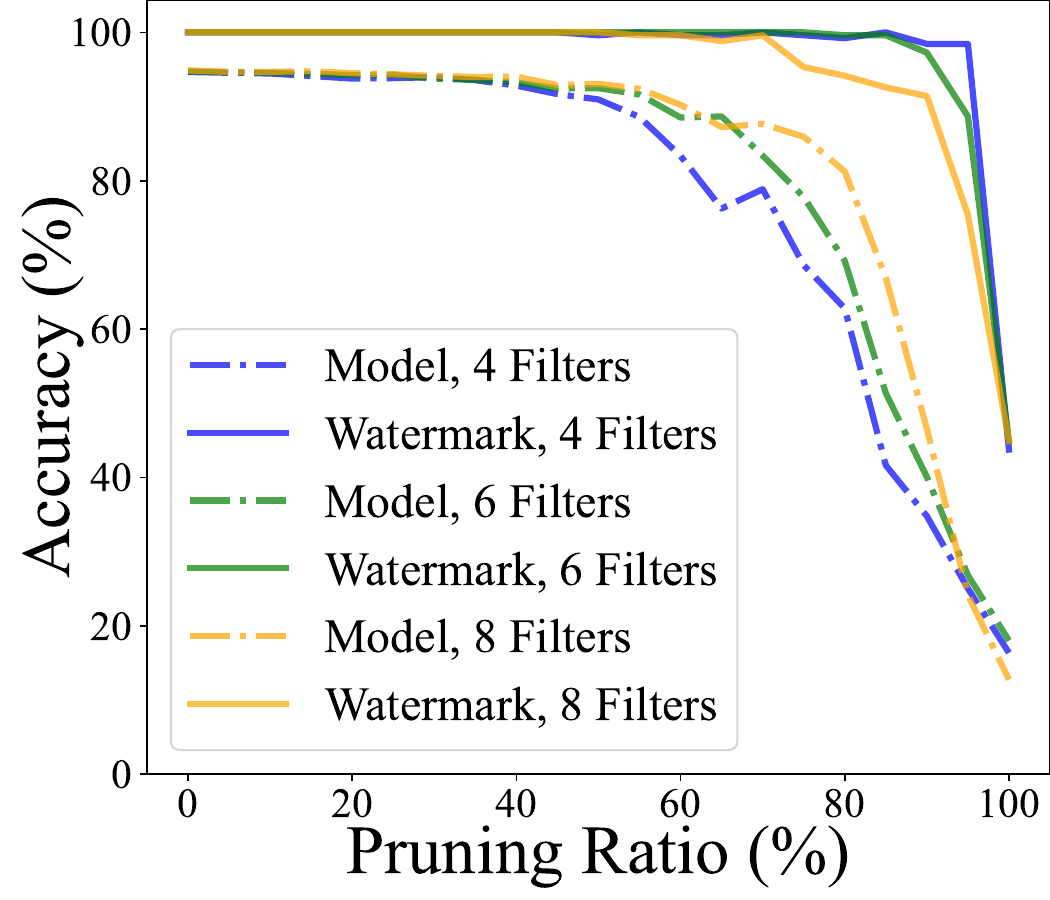}}
}
\subfloat[CIFAR-100] 
{
{\includegraphics[width=0.26\columnwidth]{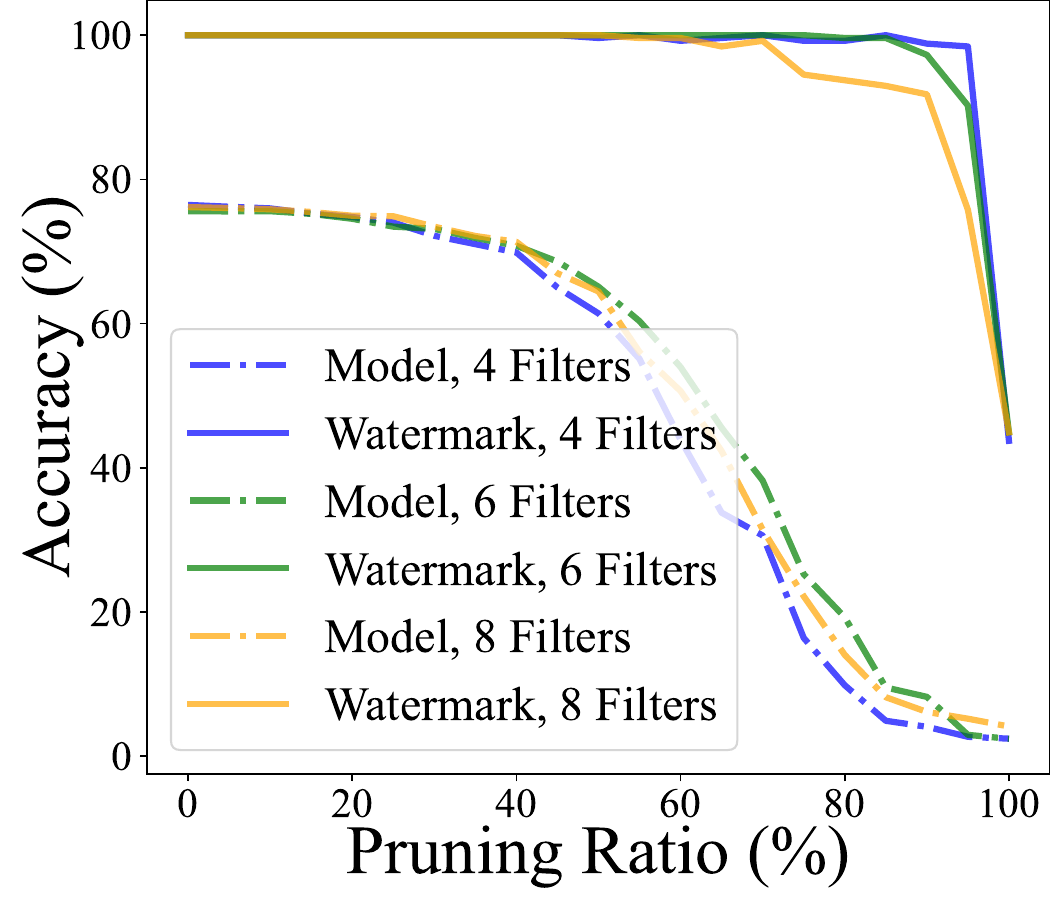}}
}
\caption{Comparison of resistance to pruning attacks with distinct filter rounds on CIFAR-10 and CIFAR-100 using ResNet-18 at various pruning ratios.}
\label{appendix_fig:PruningAttackMultiFilter}
\end{figure}

\cref{appendix_fig:PruningAttackMultiFilter} shows the results of pruning attacks on CIFAR-10 and CIFAR-100 using ResNet-18 across different filtering rounds. As can be seen, as the number of filtering rounds increases, the robustness of NeuralMark in resisting pruning attacks exhibits a slight decline. One potential reason is that increasing the number of filter rounds reduces the number of filtered parameters, leading to a smaller average pooling window size, which affects the robustness against pruning attacks to some extent.

\end{document}